\documentclass[%
  reprint,
  aps,
  prb,
  superscriptaddress,
  longbibliography
]{revtex4-2}

\usepackage{amsmath,amssymb}
\usepackage{graphicx}
\usepackage[colorlinks=true, linkcolor=blue, urlcolor=blue]{hyperref}
\usepackage{enumitem}
\usepackage{placeins}
\usepackage{booktabs}
\usepackage{siunitx}
\usepackage{orcidlink}
\usepackage{algorithm}
\usepackage{algpseudocode}
\usepackage{capt-of}

\DeclareGraphicsExtensions{.pdf,.png,.jpg}

\begin{document}

\title{Low-noise Pauli-consistent ensemble Monte Carlo for graphene\\ with electron--electron scattering}
\author{Tigran Zalinyan\,\orcidlink{0009-0008-2304-1637}}
\thanks{Corresponding author: \href{mailto:tigran.zalinyan@ysu.am}{tigran.zalinyan@ysu.am}}
\affiliation{Research Institute of Physics, Yerevan State University, Yerevan, Armenia}
\author{Giovanni Nastasi\,\orcidlink{0000-0002-6374-8441}}
\email{giovanni.nastasi@unikore.it}
\affiliation{Department of Engineering and Architecture, University of Enna ``Kore'', Enna, Italy}

\begin{abstract}
We investigate Pauli-consistent ensemble Monte Carlo simulations of graphene with explicit intraband electron--electron scattering. To reduce the cost of electron--electron proposal-rate evaluation, we introduce a sampled-partner approximation that replaces the full partner-cell sum by uniform sampling from the instantaneous ensemble, while leaving the event-level collision step unchanged. Comparison with the full-sum reference shows close agreement together with a substantial reduction in computational cost, enabling large-ensemble low-noise simulations. In this regime, systematic oscillatory components become clearly resolved in ensemble-averaged time traces. We show that these oscillations are numerical and originate from deterministic drift on the discretized momentum-space grid. We also discuss a procedure for reducing their impact in recorded observables without modifying the underlying Monte Carlo dynamics.
\end{abstract}

\maketitle

\section{Introduction}
\label{sec:secI}

Monte Carlo (MC) methods, since their introduction in the pioneering work of Metropolis and Ulam \cite{MetropolisUlam1949}, have become a powerful tool for studying systems that are difficult to treat analytically or by direct deterministic computation. In semiconductor physics, early applications to hot-carrier transport appeared as early as the 1960s \cite{kurosawa1966monte}, and MC methods have since become standard tools for studying carrier transport in semiconductors and for semiconductor device simulation \cite{JacoboniLugli1989,Jacoboni2010}.

The isolation of graphene marked a major step in the development of modern two-dimensional materials research, and its unusual electronic properties have made it a central platform for transport studies \cite{Novoselov2004,CastroNeto2009}. In graphene and related materials, transport often takes place under degenerate conditions, so the Pauli principle and final-state availability must be treated consistently in time-dependent simulations. Semiclassical transport in graphene is commonly studied through numerical solutions of the Boltzmann transport equation (BTE), using both deterministic and stochastic approaches \cite{DasSarma2011}. Among deterministic approaches, discontinuous Galerkin (DG) methods have been applied, for example, to unipolar \cite{Coco2017} and bipolar \cite{MajoranaNastasiRomano2019} graphene transport. Stochastic approaches include standard ensemble Monte Carlo (SEMC) \cite{Li2010,Fang2011} and direct simulation Monte Carlo (DSMC) methods \cite{Romano2015,Nastasi2022}. However, the consistent inclusion of Pauli blocking in Monte Carlo transport under degenerate conditions is nontrivial and has long been recognized as a methodological difficulty \cite{LugliFerry1985,Tadyszak1996,BorowikThobel1998,BorowikAdamowicz2005}. If Pauli exclusion is enforced only during collisions, unphysical occupations can still arise in the simulated distribution \cite{Tadyszak1996}. For graphene, this difficulty is addressed by the Pauli-consistent `new' ensemble Monte Carlo (NEMC) method, which enforces the occupancy constraint during both drift and collisions \cite{Romano2015,Coco2017,Coco2021}.

Electron--electron (e--e) scattering plays an important role in graphene transport and nonequilibrium dynamics \cite{Kotov2012}. It contributes to rapid carrier thermalization \cite{Dawlaty2008}, depends on screening by the surrounding environment \cite{Kim2020}, and can affect transport quantities such as the drift velocity, as shown in rate-based Monte Carlo studies \cite{Li2010,Fang2011} and in Pauli-consistent formulations \cite{Nastasi2022}. In practical calculations, the screened Coulomb interaction is commonly described within the random-phase approximation (RPA) \cite{Hwang2007}. However, including e--e scattering in a fully Pauli-consistent ensemble method is computationally demanding, because the e--e collision operator depends on the evolving distribution, and each event involves two-electron kinematics together with exact energy and momentum conservation \cite{Nastasi2022}. In particular, explicit full-sum evaluation of the e--e proposal rate can become the main runtime cost and thus limit the ensemble sizes that can be used in practice.

A further motivation for the present work is the appearance of reproducible oscillatory components in ensemble-averaged NEMC time traces. In some cases these features can already be weakly visible at moderate ensemble size, but they are not clearly resolved there because they remain partly masked by Monte Carlo noise. Their systematic study therefore requires access to the low-noise regime, where statistical fluctuations fall below the oscillation amplitude. Earlier NEMC studies of graphene were concerned mainly with the formulation and implementation of the method \cite{Romano2015,Coco2021,Nastasi2022}, while the present work focuses on the low-noise regime in which these oscillatory components can be studied in detail.

The aims of this work are to make Pauli-consistent NEMC with explicit intraband e--e scattering practical for the large ensemble sizes needed for low-noise simulations, while leaving the event-level collision step unchanged, and to use the resulting access to the low-noise regime to investigate the oscillatory components in NEMC observables. To this end, we introduce a sampled-partner approximation for the e--e proposal-rate evaluation. Random sampling of collision partners has long been used in Monte Carlo treatments of carrier--carrier scattering \cite{Brunetti1985,GoodnickLugli1988,MoskoMoskova1991}. We use this idea specifically to approximate the occupation-weighted e--e proposal-rate evaluation: the full partner-cell sum is replaced by uniform sampling of partner particles from the instantaneous ensemble. This changes only the proposal-rate evaluation, while energy--momentum conservation and Pauli blocking remain enforced for each candidate e--e event.
We validate the sampled-partner approach against the full-sum baseline, show its computational advantage, identify the numerical origin of the oscillatory components, and examine an analysis-level harmonic-subtraction procedure applied to the recorded observables.

The remainder of this paper is organized as follows. Section~\ref{sec:secII} introduces the physical model, the Pauli-consistent NEMC framework, and the sampled-partner approximation. Section~\ref{sec:secIII} presents the validation of the sampled-partner approach, its computational advantage, and the resulting access to the low-noise regime. Section~\ref{sec:secIV} studies the oscillatory components and analyzes their numerical origin.
Section~\ref{sec:secV} discusses suppression strategies and analysis-level harmonic subtraction applied to the recorded observables.
Section~\ref{sec:secVI} concludes the paper. Additional model details and supporting numerical tests are presented in the appendices.

\section{Methods}
\label{sec:secII}

\subsection{Physical model}
\label{subsec:IIA}

We consider spatially homogeneous electron transport in suspended monolayer graphene under a constant in-plane electric field $\mathbf{E}$, in an electron-doped unipolar regime with conduction-band electrons only. Accordingly, valence-band holes, interband processes such as impact ionization, and electron--hole scattering are not included. This homogeneous setting provides a controlled benchmark for studying the NEMC method and its numerical properties.

In the Dirac approximation, the energy of a conduction-band electron with wave vector
$\mathbf{k}\in\mathbb{R}^2$ is
\begin{equation}
\varepsilon(\mathbf{k})=\hbar v_F |\mathbf{k}|,
\label{eq:dirac_dispersion}
\end{equation}
where $\hbar$ is the reduced Planck constant and $v_F$ is the Fermi velocity. The corresponding
group velocity is
\begin{equation}
\mathbf{v}(\mathbf{k})=\frac{1}{\hbar}\nabla_{\mathbf{k}}\varepsilon(\mathbf{k})
= v_F\,\frac{\mathbf{k}}{|\mathbf{k}|}.
\end{equation}

The semiclassical Boltzmann transport equation for the electron distribution function
$f(t,\mathbf{k})$ reads
\begin{equation}
\frac{\partial f(t,\mathbf{k})}{\partial t}
-\frac{e}{\hbar}\,\mathbf{E}\cdot\nabla_{\mathbf{k}} f(t,\mathbf{k})
=
Q[f](t,\mathbf{k}),
\label{eq:BTE}
\end{equation}
where $e$ is the elementary charge and $Q[f](t,\mathbf{k})$ is the collision term. It consists
of electron--phonon and electron--electron contributions,
\begin{equation}
Q[f]=Q_{ep}[f]+Q_{ee}[f],
\end{equation}
where $Q_{ep}[f]$ and $Q_{ee}[f]$ are the electron--phonon and electron--electron collision operators, respectively.

For electron--phonon scattering, the collision operator is
\begin{equation}
\begin{aligned}
Q_{ep}[f](t,\mathbf{k})
=
\int_{\mathbb{R}^2}
\Big[
&\,S_{ep}(\mathbf{k}',\mathbf{k})\,f(t,\mathbf{k}')(1-f(t,\mathbf{k}))
\\
-\;&S_{ep}(\mathbf{k},\mathbf{k}')\,f(t,\mathbf{k})(1-f(t,\mathbf{k}'))
\Big]
\,d\mathbf{k}',
\end{aligned}
\label{eq:Qep}
\end{equation}
where $S_{ep}(\mathbf{k},\mathbf{k}')$ is the transition probability per unit time from
$\mathbf{k}$ to $\mathbf{k}'$.

For electron--electron scattering, the collision operator is
\begin{equation}
\begin{split}
&Q_{ee}[f](t,\mathbf{k}_1)
= \int_{\mathbb{R}^2}\!\!\int_{\mathbb{R}^2}\!\!\int_{\mathbb{R}^2}
\Big[
S_{ee}(\mathbf{k}_1',\mathbf{k}_2',\mathbf{k}_1,\mathbf{k}_2)\,
\\
&\qquad \times f(t,\mathbf{k}_1')\,f(t,\mathbf{k}_2')\,(1-f(t,\mathbf{k}_1))(1-f(t,\mathbf{k}_2))
\\
&
\qquad - S_{ee}(\mathbf{k}_1,\mathbf{k}_2,\mathbf{k}_1',\mathbf{k}_2')\,f(t,\mathbf{k}_1)\,f(t,\mathbf{k}_2)\,
\\
& \qquad  \times
(1-f(t,\mathbf{k}_1'))(1-f(t,\mathbf{k}_2'))
\Big]d\mathbf{k}_2\,d\mathbf{k}_1'\,d\mathbf{k}_2' ,
\end{split}
\label{eq:Qee}
\end{equation}
where $S_{ee}(\mathbf{k}_1,\mathbf{k}_2,\mathbf{k}_1',\mathbf{k}_2')$ is the transition
probability per unit time from the initial two-electron state
$(\mathbf{k}_1,\mathbf{k}_2)$ to the final state
$(\mathbf{k}_1',\mathbf{k}_2')$.

At $t=0$, the distribution function is the Fermi--Dirac distribution at temperature $T$
and Fermi energy $\varepsilon_F$,
\begin{equation}
f(0,\mathbf{k})=
\frac{1}{1+\exp\!\Big(\dfrac{\varepsilon(\mathbf{k})-\varepsilon_F}{k_B T}\Big)},
\label{eq:FD_init}
\end{equation}
where $k_B$ is the Boltzmann constant.

The areal carrier density, mean energy, and mean velocity are
\begin{equation}
\rho(t)=\frac{g_s g_v}{(2\pi)^2}\int_{\mathbb{R}^2} f(t,\mathbf{k})\,d\mathbf{k},
\label{eq:rho_def}
\end{equation}
\begin{equation}
\langle \varepsilon \rangle(t)=
\frac{1}{\rho(t)}\frac{g_s g_v}{(2\pi)^2}
\int_{\mathbb{R}^2} \varepsilon(\mathbf{k})\,f(t,\mathbf{k})\,d\mathbf{k},
\label{eq:mean_eps}
\end{equation}
\begin{equation}
\langle \mathbf{v} \rangle(t)=
\frac{1}{\rho(t)}\frac{g_s g_v}{(2\pi)^2}
\int_{\mathbb{R}^2} \mathbf{v}(\mathbf{k})\,f(t,\mathbf{k})\,d\mathbf{k},
\label{eq:mean_v}
\end{equation}
where $g_s=2$ and $g_v=2$ are the spin and valley degeneracies. In this spatially homogeneous unipolar model, the carrier density is conserved, so $\rho(t)=\rho(0)$.

Only normal electron--phonon processes are included; Umklapp processes are neglected.
The model includes intravalley acoustic scattering, intravalley optical scattering, and intervalley phonon
scattering. The corresponding rate expressions and parameter values are given in the Appendix~\ref{app:eph}.

The phonon occupations are fixed at their equilibrium Bose values at $T=300$~K, so hot-phonon dynamics and lattice heating are not included. Their effects on transport in suspended graphene have been studied using the DSMC method \cite{CocoRomano2018}, while strong carrier heating under high bias has been observed experimentally (see, e.g., Ref.~\cite{Kim2015}). The higher-field calculations in this work are used mainly as numerical tests of the NEMC method, not as quantitative predictions of high-field graphene transport.

For electron--electron scattering, we consider intraband binary collisions mediated by a screened Coulomb interaction. Screening is described by the equilibrium static RPA dielectric function \cite{Hwang2007}. In the Monte Carlo formulation used below, the e--e proposal rate is constructed from the loss part of the collision operator, while final-state Pauli blocking is enforced by the event-level acceptance test \cite{Nastasi2022}. The proposal rate for an electron in state $\mathbf{k}_1$ is
\begin{equation}
\lambda_{ee}(\mathbf{k}_1)=\sum_{\mathbf{k}_2} f(\mathbf{k}_2)\sum_{\mathbf{k}_1',\mathbf{k}_2'}
S_{ee}(\mathbf{k}_1,\mathbf{k}_2,\mathbf{k}_1',\mathbf{k}_2'),
\end{equation}
subject to momentum and energy conservation,
\begin{equation}
\mathbf{k}_1+\mathbf{k}_2=\mathbf{k}_1'+\mathbf{k}_2', \qquad
\varepsilon(\mathbf{k}_1)+\varepsilon(\mathbf{k}_2)=
\varepsilon(\mathbf{k}_1')+\varepsilon(\mathbf{k}_2').
\label{eq:cons}
\end{equation}
With the Dirac dispersion, the second condition is equivalent to
$|\mathbf{k}_1|+|\mathbf{k}_2|=|\mathbf{k}_1'|+|\mathbf{k}_2'|$.
According to the Fermi golden rule \cite{Li2010},
\begin{equation}
\begin{split}
& S_{ee}(\mathbf{k}_1,\mathbf{k}_2,\mathbf{k}_1',\mathbf{k}_2') \\
& =
\frac{2\pi}{\hbar}|M|^2\,
\delta\!\Big(\varepsilon(\mathbf{k}_1')+\varepsilon(\mathbf{k}_2')-\varepsilon(\mathbf{k}_1)-\varepsilon(\mathbf{k}_2)\Big).
\end{split}
\end{equation}
The interaction matrix element is
\begin{equation}
|M|^2=\frac{1}{2}\Big[\,|V(q)|^2+|V(q')|^2 - V(q)V(q')\,\Big],
\end{equation}
where $V(q)$ and $V(q')$ are the Coulomb potentials between pairs of electrons
\begin{equation}
V(q)=\frac{2\pi e^2}{\epsilon(q)\,q\,A}\,
\frac{1+\cos(\phi_{\mathbf{k}_1,\mathbf{k}_1'})}{2}\,
\frac{1+\cos(\phi_{\mathbf{k}_2,\mathbf{k}_2'})}{2},
\end{equation}
\begin{equation}
V(q')=\frac{2\pi e^2}{\epsilon(q')\,q'\,A}\,
\frac{1+\cos(\phi_{\mathbf{k}_1,\mathbf{k}_2'})}{2}\,
\frac{1+\cos(\phi_{\mathbf{k}_2,\mathbf{k}_1'})}{2},
\end{equation}
with $q=|\mathbf{k}_1-\mathbf{k}_1'|$ and $q'=|\mathbf{k}_1-\mathbf{k}_2'|$. Here $A$ is the graphene sheet area.

Screening enters via the dielectric function $\epsilon(q)$,
\begin{equation}
\epsilon(q)\,q = q + C_\epsilon\,\Pi(q).
\end{equation}
Here
\begin{equation}
C_\epsilon=\frac{r_s k_F}{2}\,(g_s g_v)^{3/2},
\end{equation}

where the Fermi wave vector is
\begin{equation}
k_F=\sqrt{\frac{4\pi\rho(0)}{g_s g_v}},
\end{equation}
and the dimensionless Wigner--Seitz radius is

\begin{equation}
r_s=\frac{e^2}{\kappa\,\hbar v_F}\sqrt{\frac{4}{g_s g_v}}.
\end{equation}
Since the carrier density is conserved, the corresponding $k_F$ remains constant during the simulation.
The background dielectric constant is taken as $\kappa=1$ for suspended graphene \cite{Reed2010}.
The static polarization function follows Ref.~\cite{Hwang2007,Kotov2012}:
\begin{equation}
\Pi(q)=
\begin{cases}
1, & q<2k_F,\\[4pt]
1+\dfrac{\pi q}{8k_F}
-\dfrac{\sqrt{q^2-4k_F^2}}{2q}\\
\qquad \qquad -\dfrac{q}{4k_F}\arcsin\!\left(\dfrac{2k_F}{q}\right),
& q\ge 2k_F .
\end{cases}
\label{eq:Pi_tilde}
\end{equation}
The equilibrium static-RPA form of the polarization function is used throughout the simulations. The momentum transfers $q$ and $q'$ are evaluated for each collision configuration, so the screened matrix element varies with the incoming and candidate outgoing states. The e--e proposal rate is calculated from the current electron distribution and evolves with it. The same screening model is used in all simulations to keep the physical collision model identical in the methodological comparisons. Neither nonequilibrium modifications of the polarization, such as drift-induced changes (see, e.g., Ref.~\cite{Sabbaghi2015}), nor frequency-dependent screening are included.

For fixed $\mathbf{k}_1$ and $\mathbf{k}_2$, the final states satisfying Eq.~\eqref{eq:cons}
form a one-dimensional manifold $\mathcal{E}$ in $\mathbf{k}$-space. For the Dirac dispersion,
this is an ellipse, parametrized by $\beta\in[0,2\pi]$~\cite{Nastasi2022}. Its semi-major axis
$a$, semi-minor axis $b$, and semi-focal distance $c$ are
\begin{equation}
2a=|\mathbf{k}_1|+|\mathbf{k}_2|,
\quad 2c=|\mathbf{k}_1+\mathbf{k}_2|,
\quad b=\sqrt{a^2-c^2},
\label{eq:abc_defs}
\end{equation}
with line element
\begin{equation}
\label{eq:d_el}
d\mathcal{E}
=
\sqrt{a^2\sin^2\beta+b^2\cos^2\beta}\,d\beta.
\end{equation}

After substitutions and standard homogenization~\cite{Jacoboni2010}, the e--e proposal rate is
\begin{equation}
\label{eq:lambda_ee_ellipse}
\lambda_{ee}(\mathbf{k}_1)
\simeq
\frac{A^2}{(2\pi)^3\,\hbar^{2} v_F}
\int_{\mathbb{R}^2}
f(\mathbf{k}_2)\,
\left[
\int_{\mathcal{E}} |M|^{2}(q,q')\, d\mathcal{E}
\right]\,
d\mathbf{k}_2 .
\end{equation}
Using Eq.~\eqref{eq:d_el}, the inner integral is
\begin{equation}
\label{eq:intE_to_beta}
\begin{aligned}
\int_{\mathcal{E}} |M|^{2}(q,q')\, d\mathcal{E}
&=
\int_{0}^{2\pi}
|M|^{2}\!\big(q(\beta),q'(\beta)\big)\,\times
\\
&\qquad
\sqrt{a^{2}\sin^{2}\beta+b^{2}\cos^{2}\beta}\, d\beta .
\end{aligned}
\end{equation}

For a detailed mathematical analysis of e--e scattering
(both intraband and interband processes), see Ref.~\cite{Nastasi2022}.

\subsection{Monte Carlo method}
\label{subsec:IIB}
\subsubsection{Pauli-consistent NEMC with full-sum evaluation of the e--e scattering rate}
\label{subsubsec:IIB1}

Transport is simulated by the Pauli-consistent `new' ensemble Monte Carlo (NEMC) method
\cite{Romano2015,Coco2021}. The method uses a drift--collision splitting on a uniform macro time grid
$t_r=r\Delta t$ and represents the distribution on an occupancy-limited $k$-space grid.

Momentum space is discretized uniformly over $\Omega_k=[-k_{\max},k_{\max}]^2$ into
$N_{k,x}\times N_{k,y}$ cells (here $N_{k,x}=N_{k,y}=N_k$) with spacing
$\Delta k=2k_{\max}/N_k$ and cell centers $\mathbf{k}_{ij}=(k_{x,i},k_{y,j})$.
The carrier distribution is represented by an ensemble of $N_p$ simulated particles. $N_p$ is not the physical number of carriers in a finite sample.
Each simulated particle follows the same drift
and scattering rules as a real carrier, while macroscopic observables are obtained from ensemble
averages.
In ensemble Monte Carlo methods, $N_p$ is a numerical parameter that controls statistical accuracy
and the resolution of the discrete occupancy field. 

At each time step, the state of the ensemble is described by the integer numbers of simulated
particles in the $k$-space cells, denoted by $\mathrm{occ}_{ij}(t)$. The initial occupancies are obtained
by discretizing Eq.~\eqref{eq:FD_init} and assigning integer occupancies $\mathrm{occ}_{ij}(0)$ for a
target ensemble size $N_{p,0}$ according to
\[
n_{ij}=N_{p,0}\,
\frac{f(0,\mathbf{k}_{ij})}{\sum_{i,j} f(0,\mathbf{k}_{ij})},
\quad
\mathrm{occ}_{ij}(0)=\mathrm{round}(n_{ij}),
\]
so that $N_p=\sum_{i,j}\mathrm{occ}_{ij}(0)$ and $M=\max_{i,j}\mathrm{occ}_{ij}(0)$. The value of $M$ is fixed throughout the simulation and defines the maximum allowed occupancy of each cell.

We then define the discrete occupation fraction
\begin{equation}
f_{ij}(t)=\frac{\mathrm{occ}_{ij}(t)}{M},
\label{eq:fij_norm}
\end{equation}
which is used in the collision step. Within each occupied cell, particle wave
vectors are initialized by uniform sampling over the cell area.

Pauli blocking is enforced for each attempted scattering event by a rejection test based on the
effective occupancies of the destination cell or cells. For an e--ph attempt from a source cell
$(i,j)$ to a destination cell $(i',j')$, we define the effective destination occupancy
\[
\mathrm{occ}^{\rm eff}_{i'j'}=
\begin{cases}
\mathrm{occ}_{i'j'}-1, & (i',j')=(i,j),\\
\mathrm{occ}_{i'j'}, & \text{otherwise,}
\end{cases}
\,\,
f^{\rm eff}_{i'j'}=\frac{\mathrm{occ}^{\rm eff}_{i'j'}}{M},
\]
draw $\eta\sim\mathcal{U}(0,1)$, and accept the event if $f^{\rm eff}_{i'j'}<\eta$
(equivalently, with probability $(1-f^{\rm eff}_{i'j'})$).

For an e--e attempt, after proposing the post-collision states $\mathbf{k}'_1,\mathbf{k}'_2$ and
mapping them to destination cells $(i'_1,j'_1)$ and $(i'_2,j'_2)$, we compute the corresponding
effective destination occupancies $\mathrm{occ}^{\rm eff}_{i'_1j'_1}$ and
$\mathrm{occ}^{\rm eff}_{i'_2j'_2}$ by subtracting the contributions of the two initial particles
whenever an initial cell coincides with a destination cell. If
$(i'_1,j'_1)=(i'_2,j'_2)$, the second test uses the updated effective occupancy to account for
the first placement. We then define
$f^{\rm eff}_{i'_1j'_1}=\mathrm{occ}^{\rm eff}_{i'_1j'_1}/M$ and
$f^{\rm eff}_{i'_2j'_2}=\mathrm{occ}^{\rm eff}_{i'_2j'_2}/M$,
draw $\eta_1,\eta_2\sim\mathcal{U}(0,1)$, and accept the event if
$f^{\rm eff}_{i'_1j'_1}<\eta_1$ and $f^{\rm eff}_{i'_2j'_2}<\eta_2$
(equivalently, with probability
$(1-f^{\rm eff}_{i'_1j'_1})(1-f^{\rm eff}_{i'_2j'_2})$).
Otherwise, the pre-collision states are retained.

The drift step follows the semiclassical streaming law
\begin{equation}
\hbar\,\frac{d\mathbf{k}}{dt}=-e\,\mathbf{E},
\end{equation}
and is implemented by a Lagrangian (co-moving) shift of the momentum grid. In this step, the
discrete occupancy field is translated rigidly in $\mathbf{k}$ space. As a result, the drift step
does not require interpolation or reconstruction of the distribution, and the occupancy constraint
enforcing the Pauli principle is preserved automatically during drift.

Collisions within each interval $[t_r,t_{r+1}]$ are treated in continuous time by a null-collision
procedure. For a particle in state $\mathbf{k}$, we define the total scattering rate
\begin{equation}
\Gamma(\mathbf{k})=\sum_{\nu}\Gamma_{\nu}(\mathbf{k}),
\label{eq:Gamma_full}
\end{equation}
where $\nu$ runs over all implemented scattering mechanisms. To generate candidate collision times,
we introduce a bounded rate $\Gamma'(\mathbf{k})=\alpha\,\Gamma(\mathbf{k})$ with $\alpha>1$.
Following Refs.~\cite{Romano2015,Coco2021,Nastasi2022}, we set $\alpha=1.1$ throughout.
The corresponding free-flight time is then sampled as
\begin{equation}
\Delta t_{\mathrm{ff}}=-\frac{\ln \eta}{\Gamma'(\mathbf{k})},
\qquad \eta\sim\mathcal{U}(0,1).
\label{eq:dt_ff}
\end{equation}
With probability $1/\alpha$, the candidate event is a real collision; otherwise it is a null collision, in which no scattering is applied and the particle state remains unchanged by the collision operator. The drift under the electric field is treated separately by the drift step on the macro time grid. Conditioned on the event being real, the scattering mechanism $\nu$ is selected with probability $\Gamma_{\nu}(\mathbf{k})/\Gamma(\mathbf{k})$, and the proposed post-collision state or states are then accepted or rejected by the event-level Pauli test.

The e--e contribution entering this mechanism selection is evaluated directly from the instantaneous
distribution on the $k$-grid by a full-sum approximation. Here
$f(\mathbf{k}_2)$ is approximated by the cell-averaged value $f_{ij}$ of the cell containing
$\mathbf{k}_2$, and the outer integral in Eq.~\eqref{eq:lambda_ee_ellipse} is approximated by the
midpoint rule. For the inner integral in Eq.~\eqref{eq:intE_to_beta}, we introduce a uniform mesh
on $[0,2\pi]$ with step $\Delta\beta = 2\pi/m_\beta$,
$0=\beta_0<\beta_1<\cdots<\beta_{m_\beta}=2\pi$, and evaluate it by the trapezoidal rule. The
resulting discrete approximation for the e--e proposal rate reads
\begin{equation}
\lambda_{ee}(\mathbf{k}_1)
\simeq
C_{ee}\sum_{i=1}^{N_k}\sum_{j=1}^{N_k} f_{ij}\,
\sum_{l=1}^{m_\beta}
\big[\mathcal{M}(\beta_{l-1})+\mathcal{M}(\beta_l)\big],
\label{eq:lambdaee_fullsum_trap}
\end{equation}
where
\begin{flalign*}
&\quad C_{ee}
\equiv
\frac{e^4\,(\Delta k)^2\,\Delta\beta}{128\,\pi\,\hbar^2\,v_F},
&\\
&\quad \mathcal{M}(\beta_l)
=
|\widetilde{M}|^2\!\big(q(\beta_l),q'(\beta_l)\big)\,
\sqrt{a^{2}\sin^{2}\beta_l+b^{2}\cos^{2}\beta_l},
&\\
&\quad |\widetilde{M}|^2\!\big(q(\beta_l),q'(\beta_l)\big)
\\
&\qquad =
|\widetilde{V}\!\big(q(\beta_l)\big)|^{2}
+
|\widetilde{V}\!\big(q'(\beta_l)\big)|^{2}
-
\widetilde{V}\!\big(q(\beta_l)\big)\,\widetilde{V}\!\big(q'(\beta_l)\big),
&\\
&\quad \widetilde{V}\!\big(q(\beta_l)\big)
=
\frac{1}{\widetilde{\epsilon}_{q(\beta_l)}}
\bigl(1+\cos(\phi_{\mathbf{k}_1,\mathbf{k}'_1})\bigr)
\bigl(1+\cos(\phi_{\mathbf{k}_2,\mathbf{k}'_2})\bigr),
&\\
&\quad \widetilde{V}\!\big(q'(\beta_l)\big)
=
\frac{1}{\widetilde{\epsilon}_{q'(\beta_l)}}
\bigl(1+\cos(\phi_{\mathbf{k}_1,\mathbf{k}'_2})\bigr)
\bigl(1+\cos(\phi_{\mathbf{k}_2,\mathbf{k}'_1})\bigr),
&&\\
&\quad \widetilde{\epsilon}_{q(\beta_l)}
\equiv
\epsilon(q(\beta_l))\,q(\beta_l)
=
q(\beta_l)+C_\epsilon\,\Pi\!\big(q(\beta_l)\big).
&&
\end{flalign*}
The prefactor $C_{ee}$ collects the Coulomb and discretization constants. The tilded quantities
$\widetilde{V}$, $\widetilde{\epsilon}$, and $|\widetilde{M}|^2$ denote the corresponding kernel terms.

When the e--e mechanism is selected, a partner electron is sampled uniformly from the current ensemble, excluding the reference particle, and $\beta\sim\mathcal{U}(0,2\pi)$ is drawn to generate an admissible post-collision pair
$(\mathbf{k}'_1(\beta),\mathbf{k}'_2(\beta))$ on the conservation manifold.
The proposed event is
then accepted or rejected by the same event-level Pauli test described above for the two destination
cells \cite{Nastasi2022}.

\subsubsection{Computational cost of explicit full-sum e--e simulations}
\label{subsubsec:IIB2}

Explicit intraband e--e scattering is substantially more expensive than e--ph scattering because the e--e proposal rate in Eq.~\eqref{eq:lambdaee_fullsum_trap} depends on the evolving distribution through the weights $f_{ij}$ and is evaluated for each queried $\mathbf{k}_1$ by a full sum over all partner cells of the $k$ grid. This evaluation also includes the $\beta$ quadrature and repeated calculations of the screened matrix element. As a result, the full-sum e--e proposal-rate evaluation makes a major contribution to the total runtime.

\begin{figure}[t]
\centering
\includegraphics[width=0.7\columnwidth]{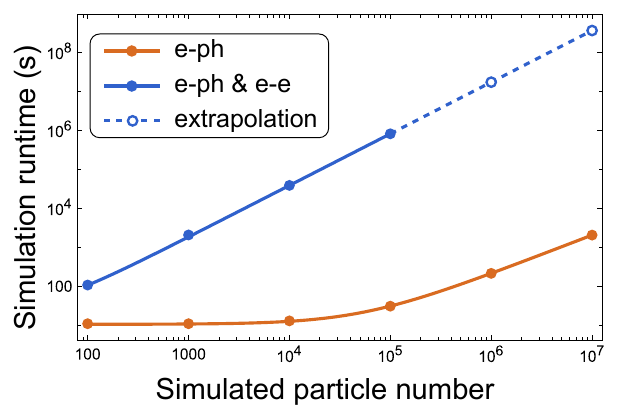}
\caption{Simulation runtime versus simulated particle number $N_p$ for NEMC simulations with only e--ph scattering and with explicit full-sum intraband e--e scattering included. Filled circles show the computed runtimes, and solid lines show empirical fits to these data. Open circles and the dashed line indicate extrapolated runtimes for the case with explicit full-sum intraband e--e scattering.}
\label{fig:1}
\end{figure}

Figure~\ref{fig:1} shows the corresponding runtimes for the parameter set
$T=300~\mathrm{K}$, $E_x=\SI{3}{kV/cm}$, $E_y=0$, $\varepsilon_F=\SI{0.15}{eV}$,
$k_{x,\max}=k_{y,\max}=\SI{3.8}{nm^{-1}}$ (corresponding to
$\varepsilon_{\max}\approx\SI{2.5}{eV}$), a uniform occupancy-limited $k$-space discretization with
$N_{k,x}\times N_{k,y}=120\times120$, a drift--collision interval $\Delta t=\SI{2.5}{fs}$,
$m_\beta=10$, and a total simulated duration $t_{\max}=\SI{5}{ps}$.
Hereafter, this parameter set
will be referred to as the baseline configuration.

The runtime data in Fig.~\ref{fig:1} show that explicit full-sum e--e scattering leads to a much larger runtime than the e--ph-only case. For the present parameter set, the runtime of the full-sum e--e simulations reaches about $10^6$~s at $N_p=10^5$, corresponding to about ten days for a single run. The extrapolated runtime is about $10^7$~s at $N_p=10^6$ and about $10^8$~s at $N_p=10^7$, corresponding to several months and several years, respectively. Thus, although the full-sum approach is suitable as a reference at moderate ensemble size, it becomes impractical for systematic simulations in the large-ensemble regime.

\subsubsection{Oscillations in NEMC time-dependent observables}
\label{subsubsec:IIB3}

\begin{figure}[!htbp]
\centering
\includegraphics[width=\columnwidth]{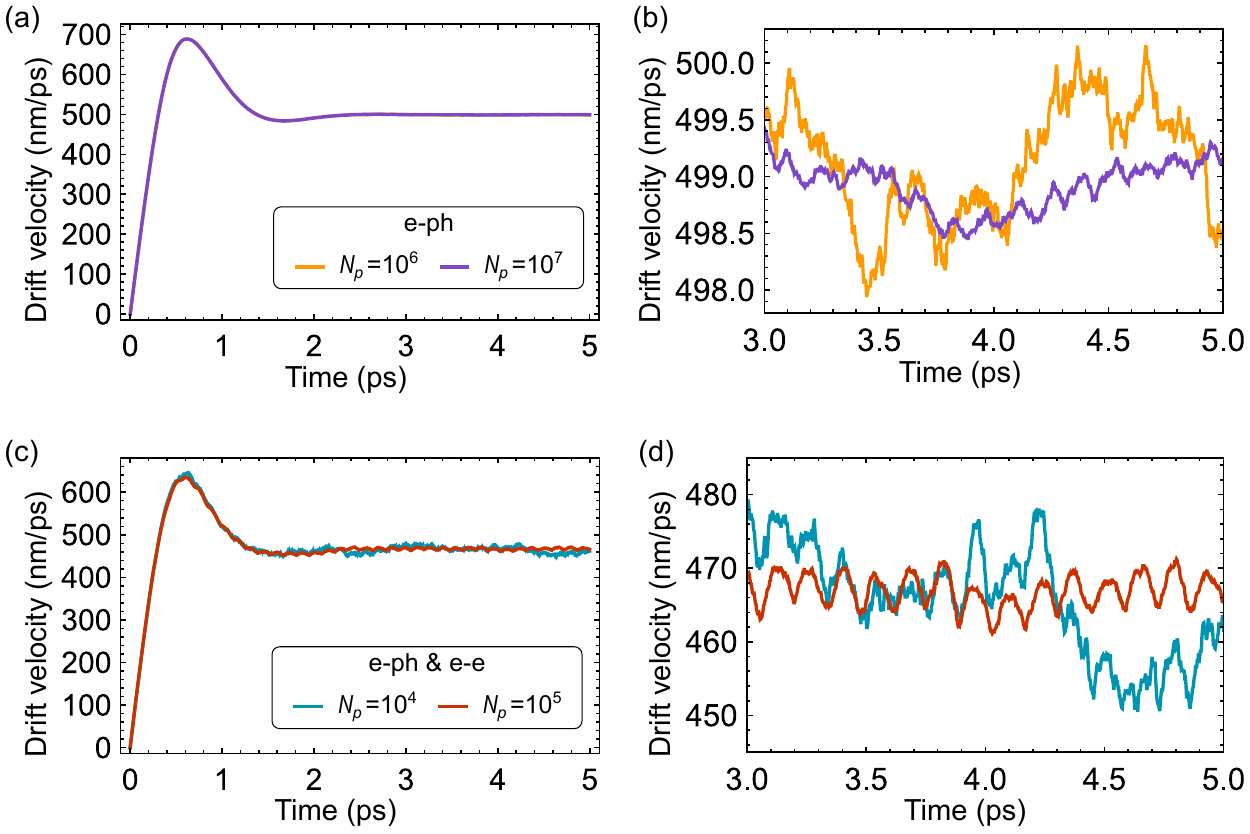}
\caption{Drift velocity for the baseline configuration. Panels (a) and (b) show the e--ph-only case for $N_p=10^6$ and $N_p=10^7$, and panels (c) and (d) show the case with explicit full-sum intraband e--e scattering for $N_p=10^4$ and $N_p=10^5$. Panels (a) and (c) show the full interval $0$--$5$ ps, while panels (b) and (d) show the zoomed interval $3$--$5$ ps.}
\label{fig:2}
\end{figure}

In the NEMC simulations considered here, ensemble-averaged observables exhibit systematic oscillations in their time dependence. As a representative example, Fig.~\ref{fig:2} shows the drift velocity.

In Figs.~\ref{fig:2}(a) and \ref{fig:2}(c), the results obtained with different ensemble sizes are difficult to distinguish over the full time interval. In the corresponding zoomed views, Figs.~\ref{fig:2}(b) and \ref{fig:2}(d), the finer structure becomes clearly visible. The oscillations are present over the full time range, but are easier to resolve in the zoomed windows.

For the e--ph-only case, shown in Figs.~\ref{fig:2}(a) and \ref{fig:2}(b), clear resolution of the oscillations requires very large ensembles. When explicit full-sum intraband e--e scattering is included, as in Figs.~\ref{fig:2}(c) and \ref{fig:2}(d), the oscillations are more pronounced and can already be resolved at smaller ensemble sizes.

These oscillations must be identified and understood, since otherwise numerical features may be mistaken for physical effects and may affect the interpretation of both transient and steady-state behavior. Their systematic study therefore requires access to a low-noise stationary regime, which in practice requires sufficiently large ensembles. Such a regime is also important for accurate evaluation of transport quantities such as the mobility, which is a key input in graphene device simulation \cite{NastasiRomano2020GFET,NastasiRomano2021EfficientGFET}. However, as shown in Sec.~\ref{subsubsec:IIB2}, simulations with explicit full-sum e--e scattering already become computationally very expensive at moderate ensemble size and are therefore impractical for systematic studies in this large-ensemble regime. This motivates the development of a more efficient approach.

\subsection{Sampled-partner method for e--e proposal-rate evaluation}
\label{subsec:IIC}

To reduce the cost of the explicit full-sum evaluation of the e--e proposal rate described in
Sec.~\ref{subsubsec:IIB1}, we introduce a sampled-partner method for e--e proposal-rate evaluation
within the same NEMC framework. The method follows the standard Monte Carlo idea of replacing an
explicit occupation-weighted sum by an average over randomly sampled partner particles from the
instantaneous ensemble \cite{JacoboniLugli1989}.

Whenever $\lambda_{ee}(\mathbf{k}_1)$ is required within the continuous-time null-collision loop, we
estimate it by drawing $N_s$ partner particles uniformly from the instantaneous ensemble, excluding
the reference particle.
Denoting the sampled partner states by $\mathbf{k}_{2,s}$, $s=1,\ldots,N_s$,
we define
\begin{equation}
\begin{aligned}
\widehat{\lambda}_{ee}(\mathbf{k}_1)
&=
\frac{N_p}{M}\,C_{ee}\,
\frac{1}{N_s}\sum_{s=1}^{N_s}
\sum_{l=1}^{m_\beta}
\Big[
\mathcal{M}\!\big(\beta_{l-1};\mathbf{k}_1,\mathbf{k}_{2,s}\big)
\\
&\qquad\qquad\qquad\qquad\qquad
+
\mathcal{M}\!\big(\beta_{l};\mathbf{k}_1,\mathbf{k}_{2,s}\big)
\Big],
\end{aligned}
\label{eq:lambdaee_sampled_partner}
\end{equation}
where $N_p$ is the current ensemble size, $M$ is the maximum cell occupancy introduced in
\eqref{eq:fij_norm}, and $C_{ee}$ and the $\beta$ mesh are the same as in
\eqref{eq:lambdaee_fullsum_trap}. Here $\mathcal{M}(\beta;\mathbf{k}_1,\mathbf{k}_2)$ denotes the
$\beta$-integrand in \eqref{eq:lambdaee_fullsum_trap} evaluated for the specific pair
$(\mathbf{k}_1,\mathbf{k}_2)$ through the ellipse parameters \eqref{eq:abc_defs} and the corresponding
$q(\beta)$ and $q'(\beta)$.

Uniform sampling over particles is equivalent to occupation-weighted sampling over Pauli cells,
because a particle drawn uniformly from the ensemble belongs to cell $(i,j)$ with probability
$\mathrm{occ}_{ij}/N_p=(M/N_p)f_{ij}$. The prefactor $N_p/M$ therefore converts the sample mean in
\eqref{eq:lambdaee_sampled_partner} into a consistent estimator of the occupation-weighted sum
underlying \eqref{eq:lambdaee_fullsum_trap}, while $1/N_s$ is the usual sample-mean normalization.

For a fixed reference state $\mathbf{k}_1$ and a fixed instantaneous ensemble, let
$Y_s(\mathbf{k}_1)$ denote the contribution of sampled partner $s$ inside the sample average in
Eq.~\eqref{eq:lambdaee_sampled_partner}. If
$\sigma_Y^2(\mathbf{k}_1)\equiv\operatorname{Var}[Y_s(\mathbf{k}_1)]$, then for independent uniform partner sampling
\[
\operatorname{Var}\!\left[\widehat{\lambda}_{ee}(\mathbf{k}_1)\right]
=
\left(\frac{N_p C_{ee}}{M}\right)^2
\frac{\sigma_Y^2(\mathbf{k}_1)}{N_s}.
\]
Thus, for a fixed ensemble state, the variance of the sampled proposal-rate estimator decreases as
$1/N_s$, while its standard deviation decreases as $1/\sqrt{N_s}$. The sample mean is unbiased for
the corresponding finite-ensemble partner average; possible accumulated effects of the stochastic
rate evaluation on the simulated observables are examined numerically in Sec.~\ref{subsec:IIIA}.

The sampled-partner method modifies only the evaluation of the e--e proposal rate entering the
channel-selection probabilities. Once an e--e event is selected, the subsequent collision update is
exactly the same as in the full-sum method: a collision partner is sampled from the ensemble, a value
$\beta\sim\mathcal{U}(0,2\pi)$ is drawn to generate an admissible post-collision pair
$(\mathbf{k}_1'(\beta),\mathbf{k}_2'(\beta))$ on the conservation manifold, and the attempt is accepted
or rejected by the same event-level Pauli test applied to both destination cells. Thus, the
approximation is confined to the stochastic evaluation of the e--e proposal intensity, whereas the
microscopic collision construction, conservation constraints, and Pauli blocking remain unchanged.

Diagnostics and validation of the sampled-partner method against the full-sum method are reported in
Sec.~\ref{sec:secIII}.


\section{Results}
\label{sec:secIII}

\subsection{Validation of the sampled-partner method}
\label{subsec:IIIA}

We begin by validating the sampled-partner method for the baseline configuration defined in Sec.~\ref{subsubsec:IIB2}. Throughout this subsection, unless otherwise specified, the simulated particle number is fixed at $N_p=10^5$, and the sampled-partner results with $N_s=100$, $10$, and $1$ are compared with the full-sum reference.

\begin{figure}[t]
  \centering
  \includegraphics[width=0.7\columnwidth]{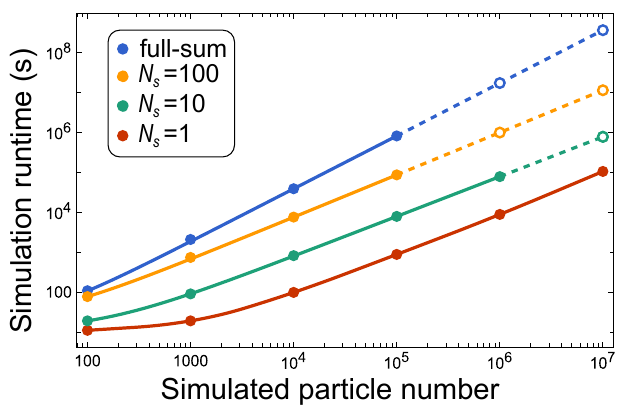}
  \caption{Simulation runtime versus simulated particle number for the baseline configuration with explicit e--e scattering. Results are shown for the full-sum reference and for the sampled-partner method with $N_s=100$, $10$, and $1$. Filled circles denote completed runs. Open circles and dashed lines indicate empirical extrapolations from the runtime data.}
  \label{fig:fig3}
\end{figure}

Fig.~\ref{fig:fig3} shows the simulation runtime as a function of the simulated particle number $N_p$. The sampled-partner method reduces the runtime strongly relative to the full-sum reference over the whole range of $N_p$, and the cost decreases further as the number of sampled partners $N_s$ is reduced.

\begin{figure}[!t]
  \centering
  \includegraphics[width=\columnwidth]{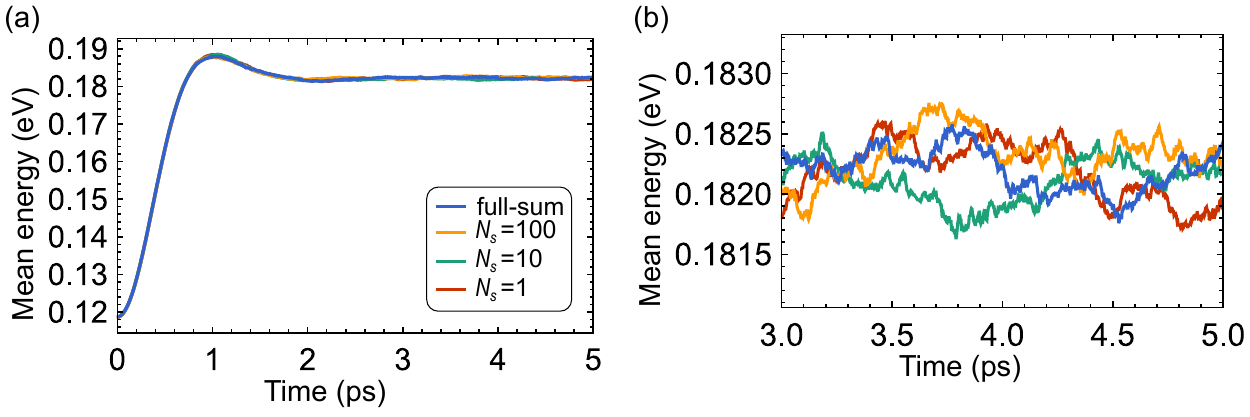}
  \caption{Mean energy for the baseline configuration at fixed $N_p=10^5$. Panel (a) shows the full time interval $0$--$5$ ps, and panel (b) shows the steady-state window. In panel (b), the curve identities are the same as in panel (a).}
  \label{fig:fig4}
\end{figure}

Figure~\ref{fig:fig4} compares the mean energy obtained with the full-sum reference and with the sampled-partner method for $N_s=100$, $10$, and $1$. Over the full simulation interval, all sampled-partner cases reproduce the same overall evolution as the full-sum result. The steady-state zoom in Fig.~\ref{fig:fig4}(b) shows that the residual fluctuations remain of the same order for all cases, with no clear systematic trend as $N_s$ is reduced.

\begin{figure}[!t]
  \centering
  \includegraphics[width=\columnwidth]{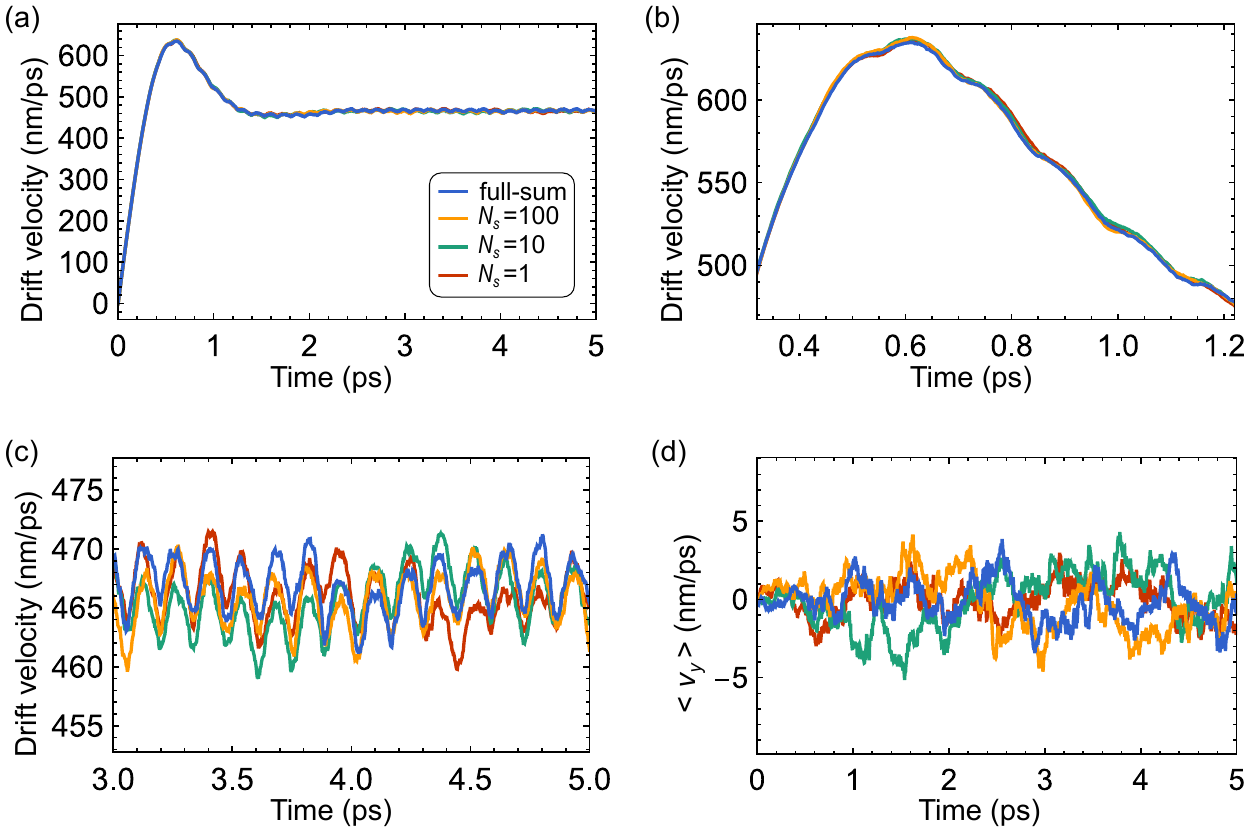}
  \caption{Velocity observables for the baseline configuration at fixed $N_p=10^5$. Panel (a) shows the drift velocity over the full time interval $0$--$5$ ps. Panel (b) shows a transient region, and panel (c) shows the steady-state window. Panel (d) shows the transverse mean velocity $\langle v_y\rangle$ over the full interval. In panels (b)--(d), the curve identities are the same as in panel (a).}
  \label{fig:fig5}
\end{figure}

Figure~\ref{fig:fig5} compares the velocity observables obtained with the full-sum reference and with the sampled-partner method. The full drift-velocity traces in Fig.~\ref{fig:fig5}(a) agree closely over the whole simulation interval. This agreement is preserved both in the transient region, as shown in Fig.~\ref{fig:fig5}(b), and in the steady-state window, as shown in Fig.~\ref{fig:fig5}(c). In particular, the steady-state zoom shows that the sampled-partner results reproduce not only the mean drift level, but also the oscillatory structure of the full-sum trace with comparable amplitude and phase. The transverse mean velocity $\langle v_y\rangle$, shown in Fig.~\ref{fig:fig5}(d), exhibits a similar fluctuation level for all cases, but no analogous oscillatory pattern.

\begin{figure}[!t]
  \centering
  \includegraphics[width=\columnwidth]{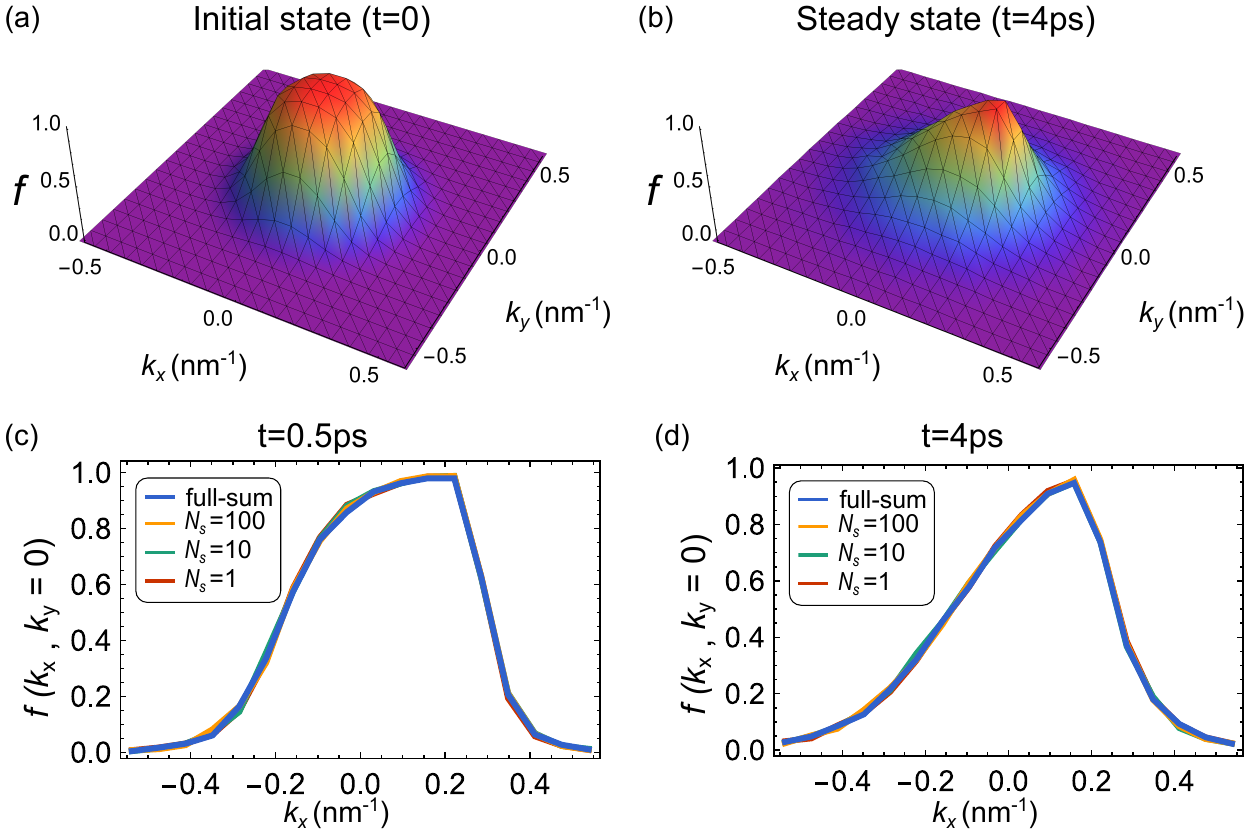}
  \caption{$k$-space distribution for the baseline configuration at fixed $N_p=10^5$. Panels (a) and (b) show the full-sum distribution at the initial time and at steady state, respectively. Panels (c) and (d) show the cut $f(k_x,k_y=0)$ at $t=0.5$ ps and $t=4$ ps, respectively, for the full-sum reference and for the sampled-partner method with $N_s=100$, $10$, and $1$.}
  \label{fig:fig6}
\end{figure}

Figure~\ref{fig:fig6} compares the $k$-space distributions obtained with the full-sum reference and with the sampled-partner method. Panels (a) and (b) illustrate the evolution of the full-sum distribution from the initial state to steady state. The corresponding cuts at $k_y=0$ are shown in panels (c) and (d) for the transient and steady-state regimes, respectively, together with the sampled-partner results for $N_s=100$, $10$, and $1$.
In both time regions, the sampled-partner cuts are in close agreement with the full-sum reference.

\begin{table*}[th]
\caption{Quantitative comparison for the baseline configuration at fixed $N_p=10^5$ over the steady window $t\in[3,5]$ ps.}
\label{tab:tab1}
\begin{ruledtabular}
\begin{tabular}{lccccc}
Method & $N_s$ & runtime (s) & $\langle \varepsilon \rangle \pm \mathrm{RMS}$ (eV) & $v_d \pm \mathrm{RMS}$ (nm/ps) & $\langle v_y\rangle \pm \mathrm{RMS}$ (nm/ps) \\
\hline
full-sum         & --  & 816398 & $0.182201 \pm 1.700\times10^{-4}$ & $466.909 \pm 2.165$ & $-0.626 \pm 1.248$ \\
sampled-partner  & 100 & 86302  & $0.182321 \pm 2.020\times10^{-4}$ & $465.778 \pm 2.218$ & $-1.212 \pm 1.199$ \\
sampled-partner  & 10  & 7859   & $0.182103 \pm 1.790\times10^{-4}$ & $465.563 \pm 2.531$ & $ 1.069 \pm 1.439$ \\
sampled-partner  & 1   & 877    & $0.182199 \pm 2.370\times10^{-4}$ & $466.007 \pm 2.325$ & $ 0.144 \pm 1.206$ \\
\end{tabular}
\end{ruledtabular}
\end{table*}

Table~\ref{tab:tab1} provides a quantitative summary of the comparison over the steady window $t\in[3,5]$ ps. For a discrete time series $x(t_r)$ sampled on the uniform simulation time grid with $\Delta t=0.0025$ ps, we retain the values with $t_r\in[3,5]$ ps and define the time average and the root-mean-square (RMS) fluctuation as
\[
\begin{aligned}
\bar{x} &= \frac{1}{N_w}\sum_{r\in W} x(t_r), \\
\mathrm{RMS}(x) &=
\left[
\frac{1}{N_w}\sum_{r\in W}\bigl(x(t_r)-\bar{x}\bigr)^2
\right]^{1/2},
\end{aligned}
\]
where $W$ is the set of retained time indices and $N_w$ is the number of samples in the window. The reported mean and RMS values are computed directly from these saved samples.
Since $E_y=0$, symmetry implies $\langle v_y\rangle=0$.
Accordingly, the values of $\langle v_y\rangle$ in Table~\ref{tab:tab1} remain of the order of their RMS fluctuations.

To quantify the agreement of the full $k$-space distributions, we define the normalized difference
\[
D_1(t)=
\frac{\sum_{ij}\left|f_{ij}^{(N_s)}(t)-f_{ij}^{(\mathrm{full})}(t)\right|}
{\sum_{ij}f_{ij}^{(\mathrm{full})}(t)},
\]
where the sums extend over all $k$-space cells. We also compare the e--e event statistics. Table~\ref{tab:rate_validation} reports the attempted and accepted e--e event rates, $\Gamma_{\mathrm{att}}$ and $\Gamma_{\mathrm{acc}}$, the acceptance probability $P_{\mathrm{acc}}$, and the mean value of $D_1$ over the steady window.

\begin{table}[!t]
\caption{Electron--electron event statistics and normalized distribution difference for the baseline configuration at fixed $N_p=10^5$ over the steady window $t\in[3,5]$ ps.}
\label{tab:rate_validation}
\begin{ruledtabular}
\begin{tabular}{lcccc}
Case &
$\Gamma_{\mathrm{att}}$ (ps$^{-1}$) &
$\Gamma_{\mathrm{acc}}$ (ps$^{-1}$) &
$P_{\mathrm{acc}}$ &
$\overline{D_1}$ (\%) \\
\hline
full-sum   & 11.18230 & 3.38316 & 0.302546 & -- \\
$N_s=100$  & 11.17368 & 3.38195 & 0.302671 & 5.4807 \\
$N_s=10$   & 11.16101 & 3.37585 & 0.302468 & 5.5446 \\
$N_s=1$    & 11.13417 & 3.36899 & 0.302581 & 5.4704 \\
\end{tabular}
\end{ruledtabular}
\end{table}

A small decrease of the event rates is observed as $N_s$ is reduced. For $N_s=1$, $\Gamma_{\mathrm{att}}$ and $\Gamma_{\mathrm{acc}}$ are about $0.43\%$ and $0.42\%$ below the full-sum values, respectively, while $P_{\mathrm{acc}}$ differs by less than $0.05\%$. The steady-state mean values in Table~\ref{tab:tab1} differ from the full-sum reference by at most $0.29\%$ for $v_d$ and $0.066\%$ for the mean energy, and these differences do not increase systematically as $N_s$ is reduced. Similarly, $\overline{D_1}$ remains close to $5.5\%$ for all three sampled-partner cases, with no systematic change as $N_s$ is reduced.

Taken together, Figs.~\ref{fig:fig4}--\ref{fig:fig6} and
Tables~\ref{tab:tab1} and \ref{tab:rate_validation} show that the
sampled-partner method remains in close agreement with the full-sum reference.
In particular, even the $N_s=1$ case reproduces the time dependence of the mean
energy and velocity observables, including the oscillatory structure, as well as
the $k$-space distribution in both the transient and steady-state regimes.
The quantitative comparisons also show no systematic change in the steady
transport observables or in the distribution-level agreement as $N_s$ is reduced.
Additional full-sum and sampled-partner comparisons for non-baseline configurations
are presented in Appendix~\ref{app:B}.

\subsection{Access to the low-noise regime}
\label{subsec:IIIB}

Having established in Sec.~\ref{subsec:IIIA} that the sampled-partner method with $N_s=1$ remains in close agreement with the full-sum reference while providing the largest runtime reduction, we now use this choice to increase the simulated particle number. Although $N_p=10^5$ is sufficient for the method comparison and for determining the steady-state mean values of the observables, $N_p=10^7$ is used to access the low-noise regime needed to resolve weak features in the time-dependent observables. Throughout this subsection, we consider the baseline configuration defined in Sec.~\ref{subsubsec:IIB2}, fix $N_s=1$, and vary $N_p$.

\begin{table*}[!t]
\caption{Quantitative comparison for the baseline configuration with fixed $N_s=1$ over the steady window $t\in[3,5]$ ps.}
\label{tab:tab2}
\begin{ruledtabular}
\begin{tabular}{lccccc}
Method & $N_p$ & runtime (s) & $\langle \varepsilon \rangle \pm \mathrm{RMS}$ (eV) & $v_d \pm \mathrm{RMS}$ (nm/ps) & $\langle v_y\rangle \pm \mathrm{RMS}$ (nm/ps) \\
\hline
sampled-partner & $10^4$  & 98     & $0.181475 \pm 4.770\times10^{-4}$ & $464.027 \pm 3.541$ & $-0.820 \pm 4.599$ \\
sampled-partner & $10^5$  & 877    & $0.182199 \pm 2.370\times10^{-4}$ & $466.007 \pm 2.325$ & $ 0.144 \pm 1.206$ \\
sampled-partner & $10^6$  & 8818   & $0.182187 \pm 5.90\times10^{-5}$ & $465.891 \pm 1.815$ & $-0.108 \pm 0.368$ \\
sampled-partner & $10^7$  & 105201 & $0.182199 \pm 1.70\times10^{-5}$ & $466.116 \pm 1.727$ & $ 0.008 \pm 0.128$ \\
\end{tabular}
\end{ruledtabular}
\end{table*}

Table~\ref{tab:tab2} summarizes the steady-window statistics for these runs, evaluated over $t\in[3,5]$ ps with the same definitions of the mean and RMS as in Sec.~\ref{subsec:IIIA}. As $N_p$ increases, the fluctuation level decreases strongly. This reduction is especially clear for the mean energy and for $\langle v_y\rangle$. By contrast, the RMS of the drift velocity decreases more slowly, since in this case it contains contributions from both Monte Carlo noise and the persistent oscillatory component.

\begin{figure}[t]
  \centering
  \includegraphics[width=\columnwidth]{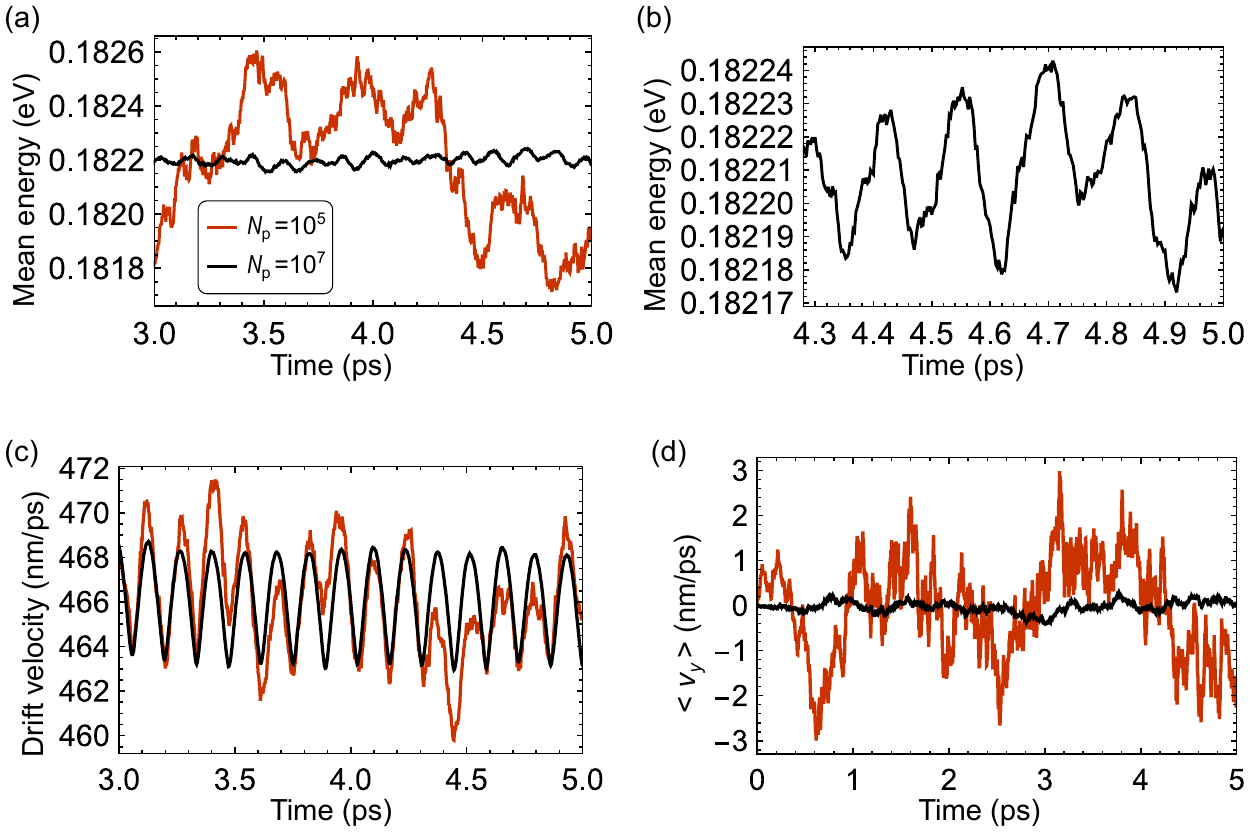}
  \caption{Access to the low-noise regime for the baseline configuration with the sampled-partner method at fixed $N_s=1$. Panel (a) shows the mean energy over the steady-state window for $N_p=10^5$ and $10^7$. Panel (b) shows a zoom of the $N_p=10^7$ mean-energy trace. Panel (c) shows the drift velocity over the steady-state window for the same two cases. Panel (d) shows the transverse mean velocity $\langle v_y\rangle$ over the full interval $0$--$5$ ps. In panels (c) and (d), the curve identities are the same as in panel (a).}
  \label{fig:fig7}
\end{figure}

The same trend is seen directly in Fig.~\ref{fig:fig7}. In the steady window, the mean-energy trace at $N_p=10^5$ is still dominated by irregular fluctuations, whereas for $N_p=10^7$ it becomes much smoother. The zoom in Fig.~\ref{fig:fig7}(b) shows that the smoother time dependence contains a regular oscillatory component. A similar effect is seen in the drift velocity: the oscillations are already visible at $N_p=10^5$, but they become much clearer and easier to resolve at $N_p=10^7$. At the same time, the transverse mean velocity $\langle v_y\rangle$ shows a strong reduction of the fluctuation level with increasing $N_p$, but no analogous oscillatory pattern.

These results show that the sampled-partner method with $N_s=1$ makes very large-ensemble simulations practical. In turn, increasing $N_p$ provides access to a low-noise regime in which the oscillations become clearly distinguishable from the statistical noise.

\section{Identification and numerical origin of oscillations}
\label{sec:secIV}

\subsection{Scaling and invariance tests}
\label{subsec:IVA}

To identify the origin of the oscillations, we first examine how their properties depend on the simulation parameters. Unless stated otherwise, this subsection uses the sampled-partner method with $N_s=1$ and the baseline configuration defined in Sec.~\ref{subsubsec:IIB2}.

In the following, we focus on the drift velocity, because the oscillations are more visible in this observable. To isolate the oscillatory part of the drift velocity, we define
\begin{equation}
\delta v_d(t)=v_d(t)-\overline{v_d},
\label{eq:deltavd_def}
\end{equation}
where $\overline{v_d}$ is the average over the plotted time window.

\begin{figure}[t]
\centering
\includegraphics[width=\columnwidth]{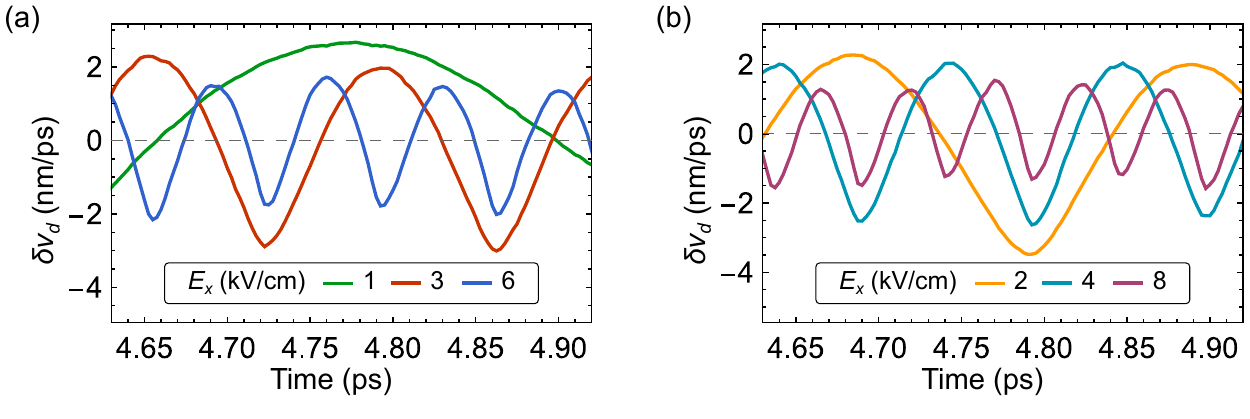}
\caption{Field dependence of $\delta v_d(t)$ at fixed discretization and
$\varepsilon_F=0.15$ eV. (a) $E_x=1$, $3$, and $6$ kV/cm. (b) $E_x=2$, $4$, and $8$ kV/cm.}
\label{fig:8}
\end{figure}

Figure~\ref{fig:8} shows $\delta v_d(t)$ for several values of $E_x$ at fixed
$\varepsilon_F=0.15$ eV and fixed $k$-space discretization. In both panels, the oscillation period
decreases as $E_x$ increases.

\begin{figure}[!t]
\centering
\includegraphics[width=\columnwidth]{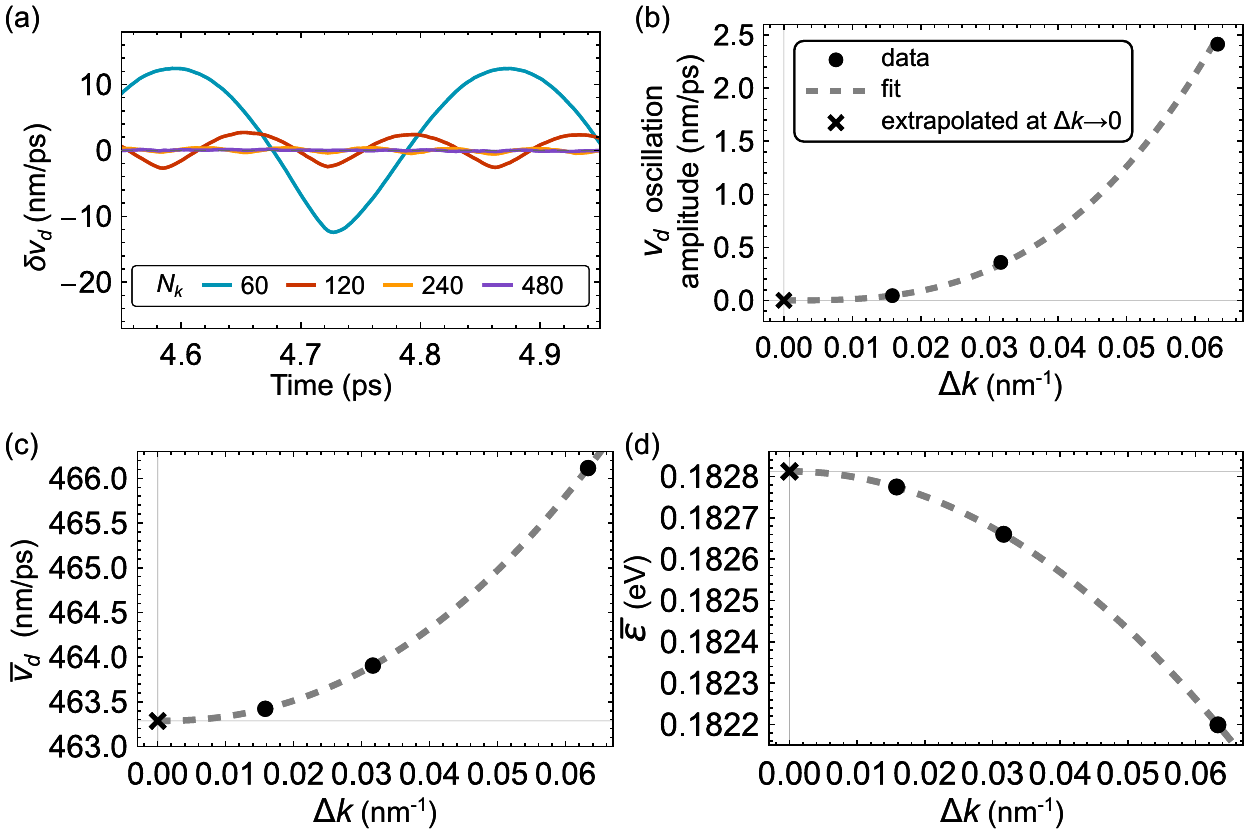}
\caption{Grid dependence for $\varepsilon_F=0.15$ eV and
$E_x=3$ kV/cm, with $N_{k_x}=N_{k_y}=N_k$.
(a) Centered drift velocity $\delta v_d(t)$ for $N_k=60$, $120$, $240$, and $480$.
(b) Oscillation amplitude versus $\Delta k$ for $N_k=120$, $240$, and $480$,
together with the power-law fit.
(c) Raw steady-state mean drift velocity $\overline{v}_d$ versus $\Delta k$.
(d) Raw steady-state mean energy $\overline{\varepsilon}$ versus $\Delta k$.
In panels (c) and (d), the three data points correspond to the same
$N_k=120$, $240$, and $480$ cases as in panel (b).
The crosses indicate the extrapolated values at $\Delta k\to0$.}
\label{fig:9}
\end{figure}

The higher-field cases included here are used as numerical tests of the field dependence of the oscillation period and are not intended as quantitative predictions of high-field graphene transport.

The dependence on the $k$-space discretization is shown in Fig.~\ref{fig:9}. Here
$\varepsilon_F=0.15$ eV and $E_x=3$ kV/cm are fixed, while the uniform grid is varied with
$N_{k_x}=N_{k_y}=N_k$. For coarse grids, the oscillatory component is large and clearly visible. As
the grid is refined, both the period and the amplitude decrease.
Figure~\ref{fig:9}(b) quantifies this decrease for the three finer grids.
The oscillation amplitudes are $2.4141$, $0.3583$, and $0.0448$ nm/ps for
$N_k=120$, $240$, and $480$, respectively.
The fitted oscillation amplitude follows approximately
$A_{\mathrm{osc}}\propto(\Delta k)^{2.88}$, showing a rapid decrease of the oscillatory component with grid refinement.
The coarse $N_k=60$ case is retained in panel~(a) to show the overall grid dependence but is not included in this fine-grid fit.

Panels~(c) and (d) show the corresponding convergence of the raw steady-state means.
Using the $N_k=120$, $240$, and $480$ results, extrapolation to $\Delta k\to0$ gives
$\overline{v}_d=463.286$ nm/ps and
$\overline{\varepsilon}=0.182812$ eV.
Relative to these extrapolated values, the baseline $N_k=120$ results differ by about
$0.61\%$ in the mean drift velocity and $0.34\%$ in the mean energy.
Thus, the raw steady-state means also approach finite values as the grid is refined.
The small grid-dependent changes in the steady-state mean values reflect the overall finite-grid
discretization, including both the oscillatory component and the grid dependence of the discrete
representation of the distribution.

\begin{figure}[!t]
\centering
\includegraphics[width=\columnwidth]{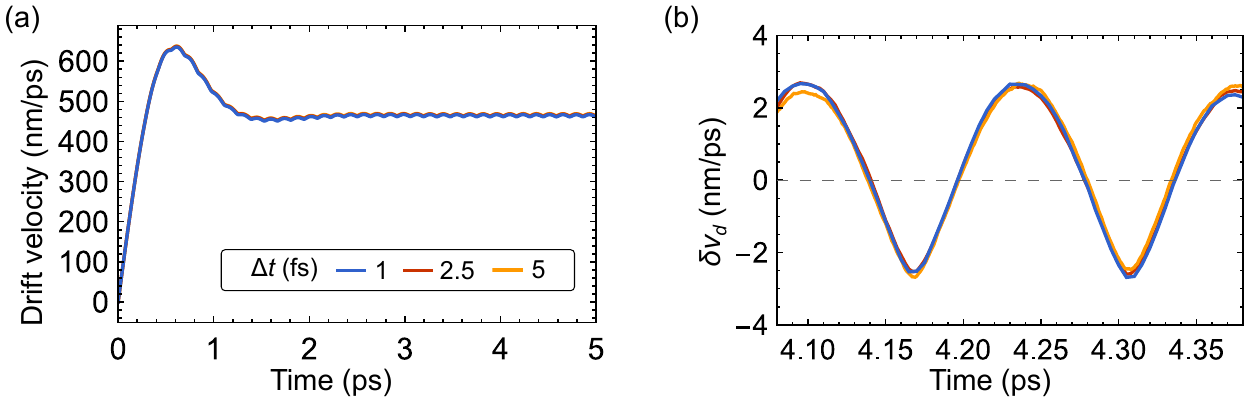}
\caption{Dependence on the macro time step for $\varepsilon_F=0.15$ eV,
$E_x=3$ kV/cm, and $N_{k_x}=N_{k_y}=120$. (a) Drift velocity $v_d(t)$ for
$\Delta t=1$, $2.5$, and $5$ fs. (b) Zoomed $\delta v_d(t)$.}
\label{fig:10}
\end{figure}

\begin{figure}[!th]
\centering
\includegraphics[width=\columnwidth]{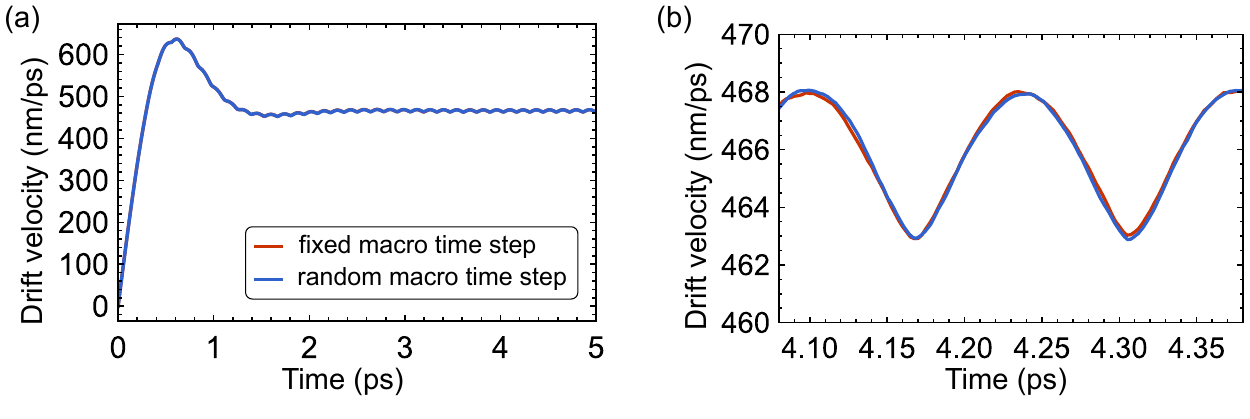}
\caption{Test with randomized macro time steps. (a) Drift velocity for fixed and randomized macro time steps. (b) Enlarged steady-state interval.}
\label{fig:random-macro-dt}
\end{figure}

Figure~\ref{fig:10} shows the dependence on the macro time step $\Delta t$ used in the
drift--collision splitting. The physical conditions and the $k$-space discretization are the same in
all three cases. Panel~(a) shows the full drift-velocity traces $v_d(t)$ for $\Delta t=1$, $2.5$,
and $5$ fs. Panel~(b) shows the corresponding $\delta v_d(t)$ in a zoomed time window. The overall
time dependence is the same, and the oscillation period does not change with $\Delta t$. The small
differences between the curves are due to residual statistical fluctuations.

To test possible synchronization with the uniform macro-time mesh, we repeated the baseline calculation using randomized macro time steps, while keeping other parameters unchanged. The reference calculation uses $\Delta t_j=\Delta t_0=2.5$~fs, whereas the randomized steps are generated as
\begin{equation*}
\Delta t_j =
\Delta t_0
\frac{q_j}{N_{\mathrm{step}}^{-1}
\sum_{\ell=1}^{N_{\mathrm{step}}}q_\ell},
\end{equation*}
where
\begin{equation*}
q_j = 1+\delta(2u_j-1),
\qquad
u_j\sim\mathcal{U}(0,1),
\end{equation*}
with $\delta=0.5$ and $N_{\mathrm{step}}=2000$. This normalization preserves $t_{\max}=N_{\mathrm{step}}\Delta t_0=5$~ps, and each $\Delta t_j$ defines the corresponding collision interval and drift increment. As shown in Fig.~\ref{fig:random-macro-dt}, randomizing the macro-time mesh leaves the oscillation essentially unchanged within the statistical noise.

\subsection{Numerical origin of the oscillations}
\label{subsec:IVB}

The results of Sec.~\ref{subsec:IVA} show that the oscillation period depends on the applied field and on the $k$-space discretization, but not on the value of the uniform macro time step, and that randomizing the macro-time mesh leaves the oscillation essentially unchanged. This points to a numerical origin connected with drift on the discretized $k$ grid and with the evaluation of final-state availability in the Pauli-consistent NEMC scheme.

The oscillations arise from the combination of two elements of the discretized Pauli-consistent NEMC scheme. The first is the occupancy-limited $k$-space discretization, which represents the distribution by a cellwise occupation field and uses this field in the event-level Pauli test. The second is the deterministic drift under a constant electric field, implemented as a global shift of the $k$ grid. The drift step moves particle states continuously in $k$ space, whereas the collision step evaluates final-state availability after assigning proposed post-collision states to destination cells of the discretized occupation field.

In the co-moving formulation, particles are stored in grid-relative coordinates $\mathbf{k}^{\mathrm{rel}}$, while the corresponding physical wave vector is
\begin{equation}
\mathbf{k}^{\mathrm{phys}}(t)=\mathbf{k}^{\mathrm{rel}}+\mathbf{k}_{\mathrm{shift}}(t),
\label{eq:k_phys_rel_IVB}
\end{equation}
where $\mathbf{k}_{\mathrm{shift}}(t)$ is determined by the drift under the applied field. During the collision step, a proposed physical post-collision state $\mathbf{k}'^{\mathrm{phys}}$ is mapped back to grid-relative coordinates and then assigned to a destination cell by the same indexing rule used to construct the occupation field.

Along the drift direction, this mapping introduces a sub-cell phase, defined by the fractional part of the drift-induced shift measured in units of the cell spacing,
\begin{equation}
\phi(t)=\left\{\frac{k_{\mathrm{shift},x}(t)}{\Delta k}\right\},
\label{eq:subcell_phase_IVB}
\end{equation}
where $\{x\}$ denotes the fractional part of $x$. Since the occupation field is constant only within each cell, the sampled destination occupancy entering the Pauli test depends on $\phi(t)$. As the drift proceeds, this phase changes continuously in time and produces a periodic variation in the acceptance statistics, which appears in ensemble-averaged observables. The initial grid phase test in Appendix~\ref{app:D_phase} supports this interpretation: changing the initial cell alignment shifts the oscillation phase, while the amplitudes remain comparable within the statistical noise.

A recurrence appears when the drift-induced shift along the field direction increases by one cell spacing, that is, when
$k_{\mathrm{shift},x}(t)\to k_{\mathrm{shift},x}(t)+\Delta k$.
Using the drift law $\hbar\,\dot{k}_x=-eE_x$, this gives the characteristic period
\begin{equation}
T_{\mathrm{grid}}=\frac{\hbar\,\Delta k}{e\,\lvert E_x\rvert}.
\label{eq:Tgrid}
\end{equation}
This period is set by the field and by the grid spacing along the drift direction.

Equivalently, introducing the dimensionless cell-crossing ratio
\[
C_k=\frac{e\lvert E_x\rvert\Delta t}{\hbar\Delta k},
\]
Eq.~\eqref{eq:Tgrid} can be written as
$T_{\mathrm{grid}}=\Delta t/C_k$.
Thus, $C_k$ can be regarded as a $k$-space Courant-like ratio for the
field-driven drift. Similar combinations of drift rate, time step, and
grid spacing arise in numerical discretizations of kinetic transport
equations~\cite{SinghBattiato2020}. Numerical recurrence associated with
finite resolution of a kinetic-space variable is also known in
Vlasov solvers~\cite{ChengGambaMorrison2013}. In the present NEMC scheme,
however, the recurrence enters specifically through the sub-cell-phase
dependence of the cellwise occupation and Pauli evaluation described above.

To test Eq.~\eqref{eq:Tgrid}, we compare the observed periods extracted from the simulations with the values given by the grid-locked formula. The comparison is shown in Table~\ref{tab:tab3} for the cases at $\varepsilon_F=0.15$ eV. Here $T_{\mathrm{obs}}$ is obtained from $\delta v_d(t)$ in the steady-state window
$t\in[2.5,5]$ ps. After subtracting the time average, we compute the discrete Fourier spectrum and define
$T_{\mathrm{obs}}$ as the inverse of the dominant nonzero frequency. Table~\ref{tab:tab3} shows close
agreement between $T_{\mathrm{obs}}$ and $T_{\mathrm{grid}}$.

\begin{table}[!th]
\caption{Observed period $T_{\mathrm{obs}}$ and grid-locked period $T_{\mathrm{grid}}$ from Eq.~\eqref{eq:Tgrid} for the cases at $\varepsilon_F=0.15$ eV.}
\label{tab:tab3}
\begin{ruledtabular}
\begin{tabular}{cccc}
$E_x$ (kV/cm) & $N_k$ & $T_{\mathrm{obs}}$ (ps) & $T_{\mathrm{grid}}$ (ps) \\
\hline
1 & 120 & 0.41708 & 0.41687 \\
2 & 120 & 0.20854 & 0.20843 \\
3 & 120 & 0.13903 & 0.13896 \\
4 & 120 & 0.10427 & 0.10422 \\
5 & 120 & 0.08342 & 0.08337 \\
6 & 120 & 0.06951 & 0.06948 \\
8 & 120 & 0.05214 & 0.05211 \\
\hline
3 &  60 & 0.27806 & 0.27791 \\
3 & 120 & 0.13903 & 0.13896 \\
3 & 240 & 0.06951 & 0.06948 \\
\end{tabular}
\end{ruledtabular}
\end{table}

Equation~\eqref{eq:Tgrid} is also consistent with the results of Fig.~\ref{fig:10}. Changing $\Delta t$ modifies the drift--collision splitting schedule, but not the drift-induced shift on the physical time axis. The cell-assignment mechanism that determines the Pauli factors therefore remains the same, and the oscillation period is unchanged.

This mechanism is not specific to e--e scattering. Oscillations are also present in the e--ph-only case. When e--e scattering is included, the oscillatory component becomes easier to resolve. First, the acceptance of an e--e event depends on the availability of two final states. Second, e--e scattering increases the number of attempted events in the energy range populated by the ensemble. As a result, the modulation appears more clearly in ensemble-averaged time traces, while the period remains unchanged and is still given by Eq.~\eqref{eq:Tgrid}.

The dependence on the observable is also consistent with the numerical mechanism described above. The oscillatory component is most visible in the drift velocity, because the grid-locked modulation acts along the field direction and therefore affects most directly the ensemble average of the velocity component parallel to the field. By contrast, the mean energy depends only on the magnitude of the wave vector, so the same phase-dependent modulation produces a weaker relative effect. No analogous oscillation is observed in $\langle v_y\rangle$, because the drift-induced recurrence is along the $x$ direction, whereas the transport setup remains symmetric in the transverse direction.

\section{Discussion: suppression and analysis-level subtraction of numerical oscillations}
\label{sec:secV}

The oscillatory components identified in Sec.~\ref{sec:secIV} are numerical rather than physical. They originate from the discretized Pauli-consistent NEMC scheme and should therefore not be interpreted as a feature of the real transport dynamics. If left untreated, they can obscure weak temporal features and complicate the interpretation of the results. It is therefore useful to examine practical strategies for suppressing their impact within the NEMC framework and for reducing their visible contribution in recorded observables through post-processing.

\subsection{Suppression strategies within the NEMC framework}
\label{subsec:V_suppression}

A natural first idea for suppressing the numerical oscillations is to refine the uniform $k$-space grid. Since the oscillations originate from the discretized drift--collision evolution, they should vanish in the formal continuum limit $\Delta k\to 0$. Consistently with the results of Sec.~\ref{subsec:IVA}, reducing the cell spacing $\Delta k$ shortens the grid-locked recurrence period and also reduces the oscillation amplitude. In practice, however, grid refinement does not eliminate the oscillations at accessible resolutions.

The basic limitation is tied to the occupancy-based Pauli representation. Pauli blocking is enforced through the integer cell occupancies $\mathrm{occ}_{ij}(t)$ and the corresponding numerical occupation fractions $f_{ij}(t)=\mathrm{occ}_{ij}(t)/M$. For fixed simulated particle number $N_p$, increasing the total number of cells reduces the typical occupancy per cell, with the rough scaling $\langle \mathrm{occ}\rangle \sim N_p/N_k^2$ for $N_{k,x}=N_{k,y}=N_k$. Beyond a certain resolution, many cells contain only a few particles, or none at all, so the cellwise occupation field becomes increasingly affected by counting noise and integer granularity. In this regime, the Pauli acceptance factors inherit this noise, and the occupation field no longer provides a stable coarse-grained representation of the underlying distribution.

In principle, one can compensate for grid refinement by increasing $N_p$ so that the per-cell statistics remain adequate as $N_k$ increases. However, maintaining a roughly fixed $\langle \mathrm{occ}\rangle$ requires $N_p$ to grow proportionally to $N_k^2$, which rapidly increases the computational cost. In addition, for the full-sum e--e treatment, grid refinement also increases the cost of the proposal-rate evaluation itself, since the sum over partner-cell occupations $f_{ij}$ is then carried out over a larger number of cells. Grid refinement is therefore an effective mitigation strategy, but not a complete practical solution.

The sampled-partner method partly alleviates this second cost increase, because the e--e proposal-rate evaluation is based on a sum over sampled partner particles rather than over all partner cells of the grid. However, it does not remove the basic occupancy-statistics limitation of the discretized Pauli representation. Thus, although grid refinement can reduce the visibility of the oscillations, it cannot be pushed arbitrarily far within the present NEMC framework.

A further possible approach is to smooth or reconstruct the cellwise Pauli factor or, more generally, to modify the event-level Pauli acceptance rule. Such modifications affect a central part of the Pauli-consistent NEMC scheme and require systematic development and validation beyond the scope of the present work. The supplementary tests presented in Appendix~\ref{app:D_smoothing} show that some reconstructions reduce the oscillatory component, but the tested schemes either violate the integer occupancy constraint or alter the steady-state transport statistics. Thus, none of them simultaneously preserves this constraint, suppresses the oscillation, and leaves the transport statistics unchanged.

\subsection{Analysis-level harmonic subtraction and validation}
\label{subsec:V_removal_validation}

The numerical origin established above naturally suggests an analysis-level treatment of the oscillatory component. This becomes possible in the low-noise regime, where Monte Carlo fluctuations are sufficiently small for the grid-locked contribution to be resolved clearly. We therefore apply harmonic subtraction as post-processing to the recorded observables. This post-processing leaves the NEMC dynamics, including the simulated ensemble evolution, collision history, and acceptance tests, unchanged.

The oscillation frequency is fixed by the grid-locked period derived in Sec.~\ref{subsec:IVB},
\begin{equation}
\omega=\frac{2\pi}{T_{\mathrm{grid}}},
\label{eq:omega_def}
\end{equation}
with $T_{\mathrm{grid}}$ given by Eq.~\eqref{eq:Tgrid}.

On a selected steady-state window, we represent an observable $y(t)$ as
\begin{equation}
y(t)=a_0+y_{\mathrm{osc}}(t)+\eta(t),
\label{eq:additive_model}
\end{equation}
where $a_0$ is a constant level, $y_{\mathrm{osc}}(t)$ is the oscillatory component, and $\eta(t)$ denotes the remaining Monte Carlo fluctuations. The steady-state window is used because, in that part of the curve, the nonoscillatory background varies only weakly and can therefore be approximated by the constant term $a_0$.

On the same window, we approximate the oscillatory component by a truncated harmonic expansion,
\begin{equation}
y_{\mathrm{osc}}^{(H)}(t)=
\sum_{h=1}^{H}\left[
b_h\sin(h\omega t)+c_h\cos(h\omega t)
\right],
\label{eq:harmonic_model_general}
\end{equation}
where the coefficients are obtained by least-squares fitting.

As discussed above, the oscillations are present over the full simulation interval, including both the transient and steady-state parts. The steady-state window is used only to determine the harmonic amplitudes and phases under conditions of minimal background variation. After these coefficients have been obtained, the fitted oscillatory component is evaluated over the full recorded time interval.

The processed trace is then defined by
\begin{equation}
y_{\mathrm{corr}}^{(H)}(t)=y(t)-y_{\mathrm{osc}}^{(H)}(t),
\label{eq:harmonic_subtract_general}
\end{equation}
so that only the oscillatory contribution is subtracted.

Higher harmonics are expected even though the recurrence period is fixed by the grid. In the present scheme, the grid-induced modulation need not be sinusoidal, because the cell-based evaluation of final-state availability and the corresponding Pauli acceptance are both discrete and nonlinear. As a result, the oscillatory component can contain higher-harmonic contributions.

\begin{figure}[!t]
\centering
\includegraphics[width=\columnwidth]{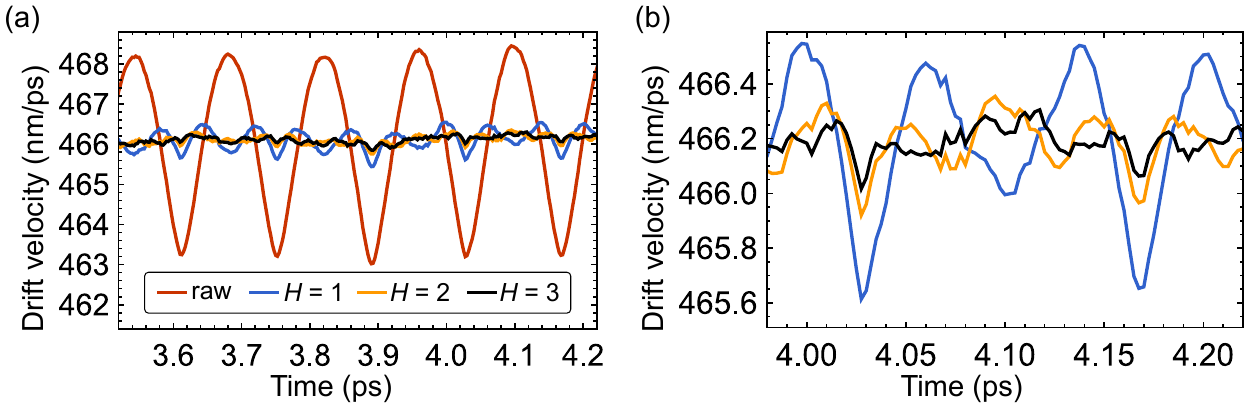}
\caption{Analysis-level harmonic subtraction for the drift velocity $v_d(t)$ in the baseline case. Panel (a) shows a steady-state segment of the raw trace together with the processed traces obtained after subtraction of harmonics up to $H=1$, $2$, and $3$, with the harmonic basis fixed by $\omega=2\pi/T_{\mathrm{grid}}$. Panel (b) shows the processed traces on an enlarged vertical scale.}
\label{fig:11}
\end{figure}

Figure~\ref{fig:11} illustrates the procedure for the drift velocity in the baseline case, obtained with the sampled-partner method at $N_s=1$ and $N_p=10^7$. Panel~(a) compares the raw trace with the traces obtained after subtracting harmonics up to $H=1$, $2$, and $3$ on the same steady-state interval. Panel~(b) shows the corresponding processed traces on an enlarged vertical scale.

The subtraction of the fundamental harmonic already reduces the dominant periodic variation, but a coherent residual remains visible. Including the second harmonic suppresses most of this residual structure, whereas the third harmonic produces only a modest further change. In the present case, these results show that a low-order harmonic expansion is sufficient to suppress the visible oscillatory component in the drift-velocity trace.

To quantify the change in the steady-state mean introduced by harmonic subtraction, we compare the mean of the raw drift-velocity trace with that of the processed trace over the same steady-state window.

Let $\overline{v}_d$ denote the mean drift velocity of the raw trace over the interval $t\in[2.5,5]$ ps, and let $\overline{v}_d^{\,(H)}$ denote the corresponding mean after subtraction of harmonics up to order $H$. We define the mean shift as
\begin{equation}
\Delta \overline{v}_d^{\,(H)}=\overline{v}_d^{\,(H)}-\overline{v}_d,
\label{eq:delta_vd_def}
\end{equation}
and define the normalized mean-shift measure
\begin{equation}
Z^{(H)}=\frac{\Delta \overline{v}_d^{\,(H)}}{\mathrm{SE}(\overline{v}_d)},
\label{eq:zscore_def}
\end{equation}
where $\mathrm{SE}(\overline{v}_d)$ is the standard error of the steady-window mean of the raw trace. The quantity $Z^{(H)}$ expresses the mean shift relative to this estimated standard error.

The results are summarized in Table~\ref{tab:V1}. For $H=1$, $2$, and $3$, $|Z^{(H)}|\ll1$ in all cases. For the baseline case, the mean shift introduced by harmonic subtraction is small compared with $\mathrm{SE}(\overline{v}_d)$.

\begin{table}[!th]
\caption{Baseline case, obtained with the sampled-partner method at $N_s=1$ and $N_p=10^7$: steady-window mean drift velocity before and after harmonic subtraction up to order $H$, evaluated over $t\in[2.5,5]$ ps.}
\label{tab:V1}
\begin{ruledtabular}
\begin{tabular}{ccccc}
$H$ & $\overline{v}_d$ (nm/ps) & $\overline{v}_d^{\,(H)}$ (nm/ps) &
$\Delta \overline{v}_d^{\,(H)}$ (nm/ps) & $Z^{(H)}$ \\
\hline
1 & 466.192388 & 466.194801 & 0.002413 & 0.012207 \\
2 & 466.192388 & 466.195109 & 0.002721 & 0.013766 \\
3 & 466.192388 & 466.195206 & 0.002818 & 0.014256 \\
\end{tabular}
\end{ruledtabular}
\end{table}

An additional non-baseline example is given in Appendix~\ref{app:C}. These results show that harmonic subtraction effectively suppresses the identified grid-locked component in the analyzed observable. As post-processing, however, it cannot undo possible effects of the phase-dependent acceptance modulation on the simulated evolution or establish that the steady-state distribution is free of grid-induced bias.


\section{Conclusions}
\label{sec:secVI}

We considered Pauli-consistent NEMC simulations of spatially homogeneous transport in electron-doped suspended monolayer graphene within a unipolar conduction-band model, with explicit intraband e--e scattering, a fixed equilibrium phonon bath, and equilibrium static-RPA screening. We introduced a sampled-partner approximation for the e--e proposal-rate evaluation. In this approximation, the full partner-cell sum is replaced by uniform sampling of partner particles from the instantaneous ensemble, while the event-level collision construction, energy--momentum conservation, and Pauli blocking are left unchanged.
Validation against the full-sum reference showed close agreement for the time-dependent observables and $k$-space distributions considered here, together with a strong reduction in computational cost.
While substantially smaller ensembles are sufficient to determine the mean transport observables, this reduction makes practical the much larger ensembles needed to reach the low-noise regime in which weak numerical oscillations can be resolved.

The low-noise regime made it possible to show that the oscillatory components observed in the NEMC dynamics are numerical and grid induced. Their period decreases with increasing electric field and with grid refinement, while remaining insensitive to changes or randomization of the macro time step. This behavior is explained by the grid-locked recurrence time $T_{\mathrm{grid}}=\hbar\,\Delta k/(e\,\lvert E_x\rvert)$, which arises from deterministic drift across the discretized $k$-space grid in the drift--collision evolution. The initial-grid-phase test and the persistence of the oscillations in e--ph-only simulations further support the conclusion that, in the present discretized NEMC implementation, the mechanism results from the coupling of this grid drift to the cellwise Pauli test. The e--e channel reinforces the modulation through the additional dependence of its proposal-rate evaluation and two final-state Pauli tests on the discretized occupation field.
Within this controlled physical setting, the higher-field calculations are used mainly as numerical tests of the NEMC method and of the grid-locked period scaling, rather than as quantitative predictions of high-field graphene transport.

We also examined practical ways to reduce the impact of this artifact. Grid refinement mitigates the oscillations but does not eliminate them at accessible resolutions and becomes increasingly costly, especially when the e--e proposal rate is evaluated by the explicit full-sum method. Local smoothing and interpolation of the Pauli factor were also tested, but reductions in the oscillatory component were accompanied either by violations of the required occupation bounds or by changes in the steady-state transport statistics. We therefore retained the original cellwise Pauli test and examined harmonic subtraction at the grid-locked frequency and its first few harmonics as an analysis-level procedure.
In the representative cases considered here, this procedure suppresses the visible oscillatory component without a statistically significant change in the corresponding steady-state mean, thereby facilitating clearer analysis and comparison of the time-dependent observables.
Because the subtraction is applied only to the recorded observables, it neither alters the simulated evolution nor removes the underlying grid-induced effect.

Taken together, these results establish the numerical origin of the weak oscillations revealed in low-noise NEMC simulations and provide an analysis-level procedure for separating the visible grid-locked contribution from the recorded observables. The sampled-partner approximation, which enables large-ensemble Pauli-consistent simulations with explicit e--e scattering, may be useful in studies where weak temporal structure or small differences in transport observables must be resolved. Although developed here for spatially homogeneous graphene transport, the computational approach may, with appropriate adaptation of the physical model, be applied to other two-dimensional or degenerate carrier systems in which Pauli blocking and carrier--carrier scattering are important. Extension to spatially nonhomogeneous transport, where temporal evolution must be treated together with real-space variations, would require further development of the method.

\section*{ACKNOWLEDGMENTS}
This research was supported by the Higher Education and Science Committee of MESCS RA
(Research Project No.~25YR-1C015).

The author G. Nastasi acknowledges the support from INdAM (GNFM) and from MUR progetto PRIN ``Transport phonema in low dimensional structures: models, simulations and theoretical aspects" CUP E53D23005900006.

Some of the calculations reported in this work were carried out on the YSU Supercomputer.

\makeatletter
\renewcommand{\appendixname}{APPENDIX}
\makeatother

\appendix

\section{ELECTRON--PHONON SCATTERING MODEL AND PARAMETERS}
\label{app:eph}
\renewcommand{\theequation}{A\arabic{equation}}
\setcounter{equation}{0}

The electron--phonon (e--ph) model is formulated for a unipolar system with conduction-band electrons only. Intravalley processes act within a single Dirac valley, whereas intervalley $K$-phonon scattering connects the $K$ and $K'$ valleys; interband processes and hole dynamics are neglected. The model includes three scattering channels: an effective intravalley acoustic channel combining the longitudinal acoustic (LA) and transverse acoustic (TA) branches, an effective intravalley optical channel combining the longitudinal optical (LO) and transverse optical (TO) branches, and an intervalley $K$-phonon channel. The effective optical channel, denoted by $O$, is characterized by phonon energy $\hbar\omega_O$ and deformation potential $D_O$ \cite{Borysenko2010,Fang2011}, while the intervalley channel is characterized by phonon energy $\hbar\omega_K$ and deformation potential $D_K$. Electrons are described in the Dirac approximation by $\varepsilon(\mathbf{k})=\hbar v_F|\mathbf{k}|$. Phonon absorption and emission are weighted by the Bose occupation
\begin{equation}
n(\omega)=\frac{1}{\exp(\hbar\omega/k_B T)-1}.
\label{eq:bose_app}
\end{equation}

For each phonon channel $\nu$, let $S_\nu(\mathbf{k},\mathbf{k}')$ denote the transition probability per unit time from $\mathbf{k}$ to $\mathbf{k}'$. The corresponding scattering rate is obtained by integrating over the final states,
\begin{equation}
\Gamma_\nu(\mathbf{k})=\int_{\mathbb{R}^2} S_\nu(\mathbf{k},\mathbf{k}')\,d\mathbf{k}'.
\label{eq:Gamma_from_S_app}
\end{equation}
In the isotropic model considered here, the scattering rates depend on $\mathbf{k}$ only through the energy, so we write $\Gamma_\nu(\mathbf{k})=\Gamma_\nu(\varepsilon)$. The total e--ph scattering rate is therefore
\begin{equation}
\Gamma_{\mathrm{ph}}(\varepsilon)=
\Gamma_{\mathrm{ac}}(\varepsilon)
+\Gamma_{O}^{\mathrm{em}}(\varepsilon)+\Gamma_{O}^{\mathrm{ab}}(\varepsilon)
+\Gamma_{K}^{\mathrm{em}}(\varepsilon)+\Gamma_{K}^{\mathrm{ab}}(\varepsilon),
\label{eq:Gamma_ph_app}
\end{equation}
where $\Gamma_{\mathrm{ac}}$ denotes the effective intravalley acoustic contribution, $\Gamma_O$ the effective intravalley optical contribution, and $\Gamma_K$ the intervalley $K$-phonon contribution \cite{Fang2011,Romano2015,Coco2021,Nastasi2022}.

The intravalley acoustic contribution is described in the equipartition approximation by the effective rate
\begin{equation}
\Gamma_{\mathrm{ac}}(\varepsilon)=
\frac{D_{\mathrm{ac}}^{2}\,k_B T}{4\,\hbar^{3}\,v_F^{2}\,\rho_m\,v_p^{2}}\;\varepsilon,
\label{eq:Gamma_ac_app}
\end{equation}
where $D_{\mathrm{ac}}$ is the acoustic deformation potential, $\rho_m$ is the areal mass density, and $v_p$ is the sound velocity \cite{Kaasbjerg2012}.

For the effective intravalley optical channel of energy $\hbar\omega_O$, the emission and absorption scattering rates are
\begin{align}
\Gamma_{O}^{\mathrm{em}}(\varepsilon)
&=
\frac{D_{O}^{2}}{\rho_m\,\omega_O\,\hbar^{2}\,v_F^{2}}\,
(\varepsilon-\hbar\omega_O)\,
\nonumber\\
&\qquad \times
(n(\omega_O)+1)\,\Theta(\varepsilon-\hbar\omega_O),
\label{eq:Gamma_O_em_app}
\end{align}
\begin{align}
\Gamma_{O}^{\mathrm{ab}}(\varepsilon)
&=
\frac{D_{O}^{2}}{\rho_m\,\omega_O\,\hbar^{2}\,v_F^{2}}\,
(\varepsilon+\hbar\omega_O)\,n(\omega_O),
\label{eq:Gamma_O_ab_app}
\end{align}
where $\Theta$ denotes the Heaviside step function.

For the intervalley $K$-phonon channel of energy $\hbar\omega_K$, the corresponding emission and absorption scattering rates are
\begin{align}
\Gamma_{K}^{\mathrm{em}}(\varepsilon)
&=
\frac{D_{K}^{2}}{\rho_m\,\omega_K\,\hbar^{2}\,v_F^{2}}\,
(\varepsilon-\hbar\omega_K)\,
\nonumber\\
&\qquad \times
(n(\omega_K)+1)\,\Theta(\varepsilon-\hbar\omega_K),
\label{eq:Gamma_K_em_app}
\end{align}
\begin{align}
\Gamma_{K}^{\mathrm{ab}}(\varepsilon)
&=
\frac{D_{K}^{2}}{\rho_m\,\omega_K\,\hbar^{2}\,v_F^{2}}\,
(\varepsilon+\hbar\omega_K)\,n(\omega_K).
\label{eq:Gamma_K_ab_app}
\end{align}

For an e--ph event at energy $\varepsilon$, the scattering channel is selected with probability proportional to the corresponding partial rate in Eq.~\eqref{eq:Gamma_ph_app}. The post-collision energy is
\begin{equation}
\varepsilon'=
\begin{cases}
\varepsilon, & \text{acoustic},\\
\varepsilon\pm\hbar\omega_O, & \text{optical absorption/emission},\\
\varepsilon\pm\hbar\omega_K, & \text{$K$ absorption/emission},
\end{cases}
\label{eq:eph_energy_update_app}
\end{equation}
with emission allowed only above threshold. The post-collision wave-vector magnitude then follows from energy conservation as $|\mathbf{k}'|=\varepsilon'/(\hbar v_F)$.

The post-collision direction is sampled from standard angular distributions for electrons with
linear dispersion. If $\theta$ is the scattering angle between the incoming and outgoing wave
vectors, the angular probability density is proportional to $\tfrac{1}{2}(1+\cos\theta)$ for
intravalley acoustic scattering and to $\tfrac{1}{2}(1-\cos\theta)$ for intervalley $K$
scattering. The effective intravalley optical scattering is taken isotropic, so the polar angle is
sampled uniformly on $[0,2\pi]$. The outgoing wave vector $\mathbf{k}'$ is then obtained from the
sampled direction and the magnitude $|\mathbf{k}'|$.

Table~\ref{tab:eph_params} lists the parameter values used in the e--ph model
\cite{Fang2011,Romano2015,Coco2021,Nastasi2022}.

\begin{table}[!th]
\caption{Parameters used in the electron--phonon model.}
\label{tab:eph_params}
\begin{ruledtabular}
\begin{tabular}{lll}
Quantity & Symbol & Value \\
\hline
Fermi velocity & $v_F$ & $1.0\times 10^{6}\ \mathrm{m/s}$ \\
Areal mass density & $\rho_m$ & $7.6\times 10^{-8}\ \mathrm{g/cm^2}$ \\
Sound velocity & $v_p$ & $2.13\times 10^{4}\ \mathrm{m/s}$ \\
Acoustic deformation potential & $D_{\mathrm{ac}}$ & $6.8\ \mathrm{eV}$ \\
Optical phonon energy & $\hbar\omega_O$ & $164.6\ \mathrm{meV}$ \\
Intervalley phonon energy & $\hbar\omega_K$ & $124\ \mathrm{meV}$ \\
Optical deformation potential & $D_O$ & $1.0\times 10^{9}\ \mathrm{eV/cm}$ \\
Intervalley deformation potential & $D_K$ & $3.5\times 10^{8}\ \mathrm{eV/cm}$ \\
\end{tabular}
\end{ruledtabular}
\end{table}

\section{ADDITIONAL FULL-SUM AND SAMPLED-PARTNER COMPARISONS}
\label{app:B}
\renewcommand{\theequation}{B\arabic{equation}}
\setcounter{equation}{0}

In Sec.~\ref{subsec:IIIA}, the sampled-partner method was compared in detail with the full-sum reference for the baseline configuration at $N_p=10^5$. Here we extend this comparison to the tested combinations of $\varepsilon_F$ and $E_x$ at $N_p=10^4$.

Table~\ref{tab:tabB1} summarizes the comparison over the steady window $t\in[3,5]$ ps for these additional cases. In all tested cases, the sampled-partner results remain close to the full-sum reference over the range $N_s=100,10,1$, while providing a substantial reduction in runtime.

\makeatletter
\setlength{\@fptop}{0pt}
\setlength{\@dblfptop}{0pt}
\makeatother
\begin{table*}[!th]
\caption{Full-sum and sampled-partner comparison for the tested combinations of $\varepsilon_F$ and $E_x$ at fixed $N_p=10^4$, evaluated over the steady window $t\in[3,5]$ ps. Here f-s and s-p denote the full-sum and sampled-partner methods, respectively.}
\label{tab:tabB1}
\begin{ruledtabular}
\begin{tabular}{cccccccc}
$\varepsilon_F$ (eV) & $E_x$ (kV/cm) & Method & $N_s$ & runtime (s) & $\langle \varepsilon \rangle \pm \mathrm{RMS}$ (eV) & $v_d \pm \mathrm{RMS}$ (nm/ps) & $\langle v_y\rangle \pm \mathrm{RMS}$ (nm/ps) \\
\hline
0.15 & 1 & f-s & --  & 23837 & $0.146103 \pm 5.6\times10^{-4}$ & $438.057 \pm 5.568$ & $-1.409 \pm 3.421$ \\
     &   & s-p & 100 & 7302  & $0.146448 \pm 6.1\times10^{-4}$ & $437.293 \pm 5.007$ & $ 0.238 \pm 2.780$ \\
     &   & s-p & 10  & 792   & $0.146568 \pm 7.8\times10^{-4}$ & $445.628 \pm 8.897$ & $-2.732 \pm 5.315$ \\
     &   & s-p & 1   & 96    & $0.146422 \pm 4.4\times10^{-4}$ & $440.008 \pm 3.808$ & $-1.512 \pm 2.912$ \\
\hline
0.15 & 3 & f-s & --  & 38789 & $0.182274 \pm 6.2\times10^{-4}$ & $465.401 \pm 7.196$ & $ 3.940 \pm 5.103$ \\
     &   & s-p & 100 & 7550  & $0.181559 \pm 6.5\times10^{-4}$ & $470.729 \pm 4.881$ & $ 0.937 \pm 4.141$ \\
     &   & s-p & 10  & 817   & $0.181649 \pm 6.4\times10^{-4}$ & $465.402 \pm 4.690$ & $-1.378 \pm 3.763$ \\
     &   & s-p & 1   & 98    & $0.181475 \pm 4.8\times10^{-4}$ & $464.027 \pm 3.541$ & $-0.820 \pm 4.599$ \\
\hline
0.15 & 5 & f-s & --  & 51987 & $0.211362 \pm 5.1\times10^{-4}$ & $463.727 \pm 5.389$ & $ 0.675 \pm 4.096$ \\
     &   & s-p & 100 & 7751  & $0.212098 \pm 1.1\times10^{-3}$ & $464.111 \pm 7.243$ & $ 0.384 \pm 4.766$ \\
     &   & s-p & 10  & 838   & $0.211222 \pm 5.7\times10^{-4}$ & $464.213 \pm 4.705$ & $-0.653 \pm 4.320$ \\
     &   & s-p & 1   & 101   & $0.212246 \pm 7.0\times10^{-4}$ & $464.022 \pm 4.957$ & $ 0.659 \pm 5.057$ \\
\hline
0.25 & 1 & f-s & --  & 25984 & $0.193979 \pm 4.7\times10^{-4}$ & $319.165 \pm 6.732$ & $-0.532 \pm 3.908$ \\
     &   & s-p & 100 & 7291  & $0.193779 \pm 3.9\times10^{-4}$ & $321.826 \pm 3.073$ & $ 3.398 \pm 2.090$ \\
     &   & s-p & 10  & 786   & $0.193872 \pm 3.4\times10^{-4}$ & $323.516 \pm 2.855$ & $ 1.565 \pm 2.447$ \\
     &   & s-p & 1   & 94    & $0.193919 \pm 2.6\times10^{-4}$ & $320.666 \pm 2.555$ & $ 2.045 \pm 1.749$ \\
\hline
0.25 & 3 & f-s & --  & 35789 & $0.218953 \pm 4.3\times10^{-4}$ & $376.084 \pm 3.163$ & $ 3.809 \pm 3.256$ \\
     &   & s-p & 100 & 7367  & $0.218949 \pm 7.0\times10^{-4}$ & $375.979 \pm 5.631$ & $ 0.315 \pm 5.896$ \\
     &   & s-p & 10  & 797   & $0.219489 \pm 3.8\times10^{-4}$ & $377.030 \pm 2.632$ & $-1.892 \pm 3.346$ \\
     &   & s-p & 1   & 97    & $0.218799 \pm 6.1\times10^{-4}$ & $372.835 \pm 3.684$ & $-0.926 \pm 4.291$ \\
\hline
0.25 & 5 & f-s & --  & 47523 & $0.242497 \pm 5.1\times10^{-4}$ & $391.016 \pm 3.524$ & $ 3.908 \pm 2.978$ \\
     &   & s-p & 100 & 7500  & $0.242068 \pm 1.1\times10^{-3}$ & $391.170 \pm 6.519$ & $-1.324 \pm 4.176$ \\
     &   & s-p & 10  & 811   & $0.242618 \pm 5.5\times10^{-4}$ & $391.128 \pm 5.306$ & $-0.640 \pm 3.129$ \\
     &   & s-p & 1   & 98    & $0.242396 \pm 1.1\times10^{-3}$ & $392.386 \pm 6.286$ & $ 2.010 \pm 4.389$ \\
\end{tabular}
\end{ruledtabular}
\end{table*}

\begin{figure}[!t]
  \centering
  \includegraphics[width=\columnwidth]{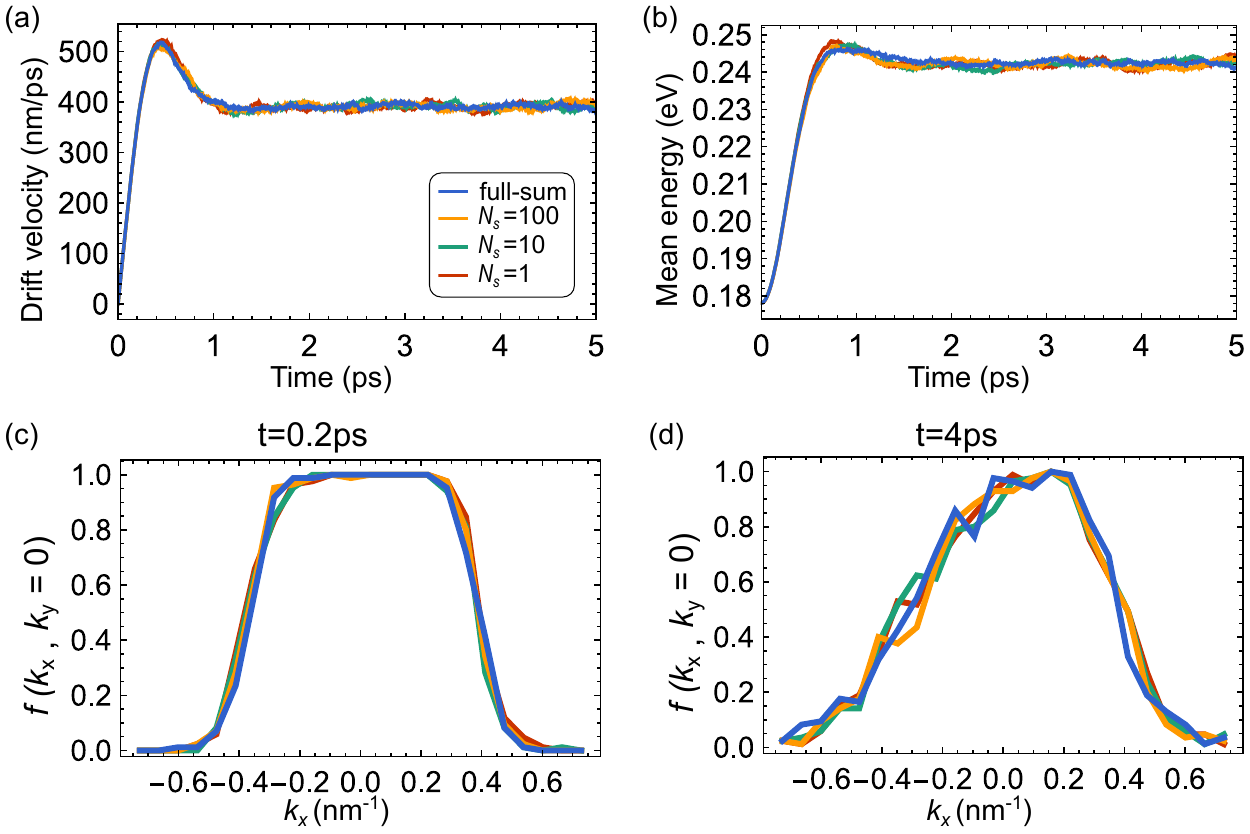}
\caption{Representative full-sum and sampled-partner comparison for a non-baseline configuration, $\varepsilon_F=0.25$ eV and $E_x=5$ kV/cm, at fixed $N_p=10^4$. Panels (a) and (b) show the drift velocity and mean energy over the full time interval. Panels (c) and (d) show the cut $f(k_x,k_y=0)$ at $t=0.2$ ps and $t=4$ ps, respectively. Results are shown for the full-sum method and for the sampled-partner method with $N_s=100$, $10$, and $1$. In panels (b)--(d), the curve identities are the same as in panel (a).}
  \label{fig:figB1}
\end{figure}

As a representative example, Fig.~\ref{fig:figB1} shows the case $\varepsilon_F=0.25$ eV and $E_x=5$ kV/cm at fixed $N_p=10^4$. Panels (a) and (b) compare the drift velocity and mean energy over the full time interval, while panels (c) and (d) compare the distribution cut $f(k_x,k_y=0)$ at $t=0.2$ ps and $t=4$ ps. The sampled-partner results remain close to the full-sum reference both for the ensemble-averaged observables and for the underlying $k$-space distribution.

\section{ANALYSIS-LEVEL HARMONIC SUBTRACTION FOR A NON-BASELINE CONFIGURATION}
\label{app:C}
\renewcommand{\theequation}{C\arabic{equation}}
\setcounter{equation}{0}

As a non-baseline example of the procedure described in Sec.~\ref{subsec:V_removal_validation}, Table~\ref{tab:tabC2} reports the case $\varepsilon_F=0.25$ eV and $E_x=3$ kV/cm, obtained with the sampled-partner method at $N_s=1$ and $N_p=10^7$.
The mean shifts satisfy $|Z^{(H)}|\ll1$ for $H=1,2,3$, indicating that the changes introduced by harmonic subtraction are small compared with $\mathrm{SE}(\overline{v}_d)$ in this case as well.

\begin{center}
\begin{minipage}{\columnwidth}
\captionof{table}{Steady-window mean drift velocity before and after harmonic subtraction up to order $H$ for the case $\varepsilon_F=0.25$ eV and $E_x=3$ kV/cm, evaluated over $t\in[2.5,5]$ ps.}
\label{tab:tabC2}
\begin{ruledtabular}
\begin{tabular}{ccccc}
$H$ & $\overline{v}_d$ (nm/ps) & $\overline{v}_d^{\,(H)}$ (nm/ps) &
$\Delta \overline{v}_d^{\,(H)}$ (nm/ps) & $Z^{(H)}$ \\
\hline
1 & 375.390570 & 375.391047 & 0.000477 & 0.011418 \\
2 & 375.390570 & 375.391112 & 0.000542 & 0.012973 \\
3 & 375.390570 & 375.391133 & 0.000563 & 0.013477 \\
\end{tabular}
\end{ruledtabular}
\end{minipage}
\end{center}

\section{ADDITIONAL TESTS OF GRID-INDUCED OSCILLATIONS}
\label{app:D}
\renewcommand{\theequation}{D\arabic{equation}}
\setcounter{equation}{0}

\subsection{Initial grid phase test}
\label{app:D_phase}

To test the dependence of the oscillation on the initial alignment between the distribution and the occupancy grid, we repeated the baseline calculation defined in Sec.~\ref{subsubsec:IIB2} for four initial grid phases. The initial grid shift is
\begin{equation}
\mathbf{k}_{\mathrm{shift}}(0)
=
\phi_0\Delta k\,\hat{\mathbf{x}},
\qquad
\phi_0\in
\left\{
0,\frac{1}{4},\frac{1}{2},\frac{3}{4}
\right\}.
\label{eq:D_phase_shift}
\end{equation}
In grid-relative coordinates, the corresponding initial distribution is
\begin{equation}
f_0^{(\phi_0)}(\mathbf{k}^{\mathrm{rel}})
=
f_{\mathrm{FD}}\!\left(
\mathbf{k}^{\mathrm{rel}}
+\phi_0\Delta k\,\hat{\mathbf{x}}
\right).
\label{eq:D_phase_initialization}
\end{equation}
Varying $\phi_0$ changes the cell alignment without changing the physical Fermi--Dirac distribution. All other numerical parameters are unchanged. The calculations include both e--ph and sampled-partner e--e scattering, with $N_s=1$ and $N_p=10^7$.

The phase mean is defined as
\begin{equation}
v_d^{\mathrm{ph-av}}(t)
=
\frac{1}{4}
\sum_{\phi_0}
v_d(t;\phi_0).
\label{eq:D_phase_average}
\end{equation}

\begin{figure}[!th]
\centering
\includegraphics[width=\columnwidth]{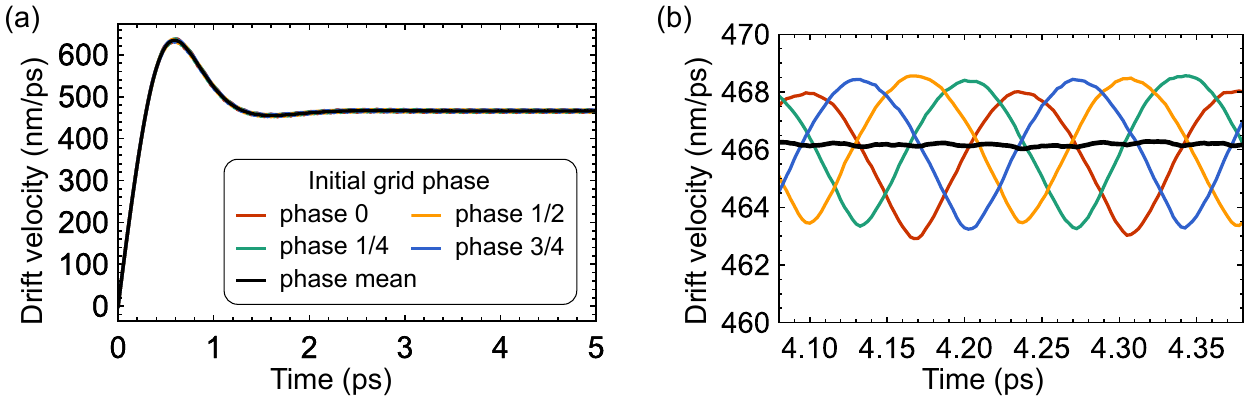}
\caption{Initial grid phase test for the baseline configuration, with $N_p=10^7$ and $N_s=1$. (a) Drift velocity over $0$--$5$ ps for the four initial grid phases and their phase mean. (b) Enlarged steady-state interval. The curve identities are the same in both panels.}
\label{fig:figD1}
\end{figure}

Figure~\ref{fig:figD1} shows that varying $\phi_0$ shifts the oscillation phase, while the differences in amplitude remain within the statistical noise. The periodic component is strongly suppressed in the phase mean, consistent with locking to the relative phase between the moving grid and the distribution.

\subsection{Smoothing and interpolation tests for the Pauli factor}
\label{app:D_smoothing}

To test whether the oscillation can be reduced by a local reconstruction of the Pauli factor, we replaced the exact cell occupation fraction $f_{ij}$ of Eq.~\eqref{eq:fij_norm} entering the event-level Pauli test with a reconstructed value $\widetilde f$. The reconstruction is evaluated from the event-adjusted field $f_{ij}^{\mathrm{eff}}$, while the stored integer occupation field, collision proposals, and integer updates remain unchanged. The tests use the baseline configuration defined in Sec.~\ref{subsubsec:IIB2}, with initial grid phase $\phi_0=0$ and the same initial ensemble and random seed.

For bilinear interpolation, let $(i_\alpha,j_\beta)$, with $\alpha,\beta\in\{0,1\}$, denote the four cell centers surrounding the proposed state $\mathbf{k}'$. With fractional coordinates $\xi,\zeta\in[0,1]$,
\begin{equation}
\begin{aligned}
\widetilde f_{\mathrm{B}}^{\mathrm{eff}}(\mathbf{k}')
={}&
\sum_{\alpha,\beta=0}^{1}
w_\alpha(\xi)w_\beta(\zeta)
f_{i_\alpha j_\beta}^{\mathrm{eff}},
\\
w_0(s)={}&1-s,
\qquad
w_1(s)=s.
\end{aligned}
\label{eq:D_bilinear}
\end{equation}

For the $3\times3$ kernel reconstruction at destination cell $D=(i',j')$,
\begin{equation}
\widetilde f_{\mathrm{K}}^{\mathrm{eff}}(D)
={}
\frac{1}{W_D}
\sum_{(a,b)\in\mathcal{N}_D}
K_{ab}f_{i'+a,j'+b}^{\mathrm{eff}} ,
\label{eq:D_kernel}
\end{equation}
where $\mathcal{N}_D$ contains the valid neighboring cells, and
\begin{equation*}
W_D={}
\sum_{(a,b)\in\mathcal{N}_D}K_{ab},
\qquad
K={}
\begin{pmatrix}
1&2&1\\
2&4&2\\
1&2&1
\end{pmatrix}.
\end{equation*}

The clipped local-linear reconstruction is
\begin{equation}
\begin{aligned}
\widetilde f_{\mathrm{L}}^{\mathrm{eff}}(\mathbf{k}')
={}&
\operatorname{clip}_{[0,1]}\!\Bigl[
f_{i'j'}^{\mathrm{eff}}
+\delta_x(D_xf^{\mathrm{eff}})_{i'j'}
\\
&+\delta_y(D_yf^{\mathrm{eff}})_{i'j'}
\Bigr].
\end{aligned}
\label{eq:D_local_linear}
\end{equation}
The quantities $\delta_x$ and $\delta_y$ are the displacements from the destination-cell center in units of $\Delta k$. Central differences are used in the interior and one-sided differences at the boundaries.

The three reconstructions, with and without an explicit integer cap, are compared with the exact-cell baseline. In the capped variants, an event is rejected if its acceptance would make the occupancy of any destination cell exceed $M$.

For the quantitative comparison, $A/A_0$ is the ratio of the fundamental-harmonic amplitude $A$ to its value $A_0$ for the exact-cell baseline. The changes $\Delta\overline{v}_d$ and $\Delta\overline{\varepsilon}$ are evaluated relative to the corresponding baseline means.

The quantity
\begin{equation}
R_{\max}
=
\max_{r,i,j}
\frac{\mathrm{occ}_{ij}(t_r)}{M}
\label{eq:D_Rmax}
\end{equation}
measures violation of the integer occupancy constraint.

\begin{figure}[!th]
\centering
\includegraphics[width=\columnwidth]{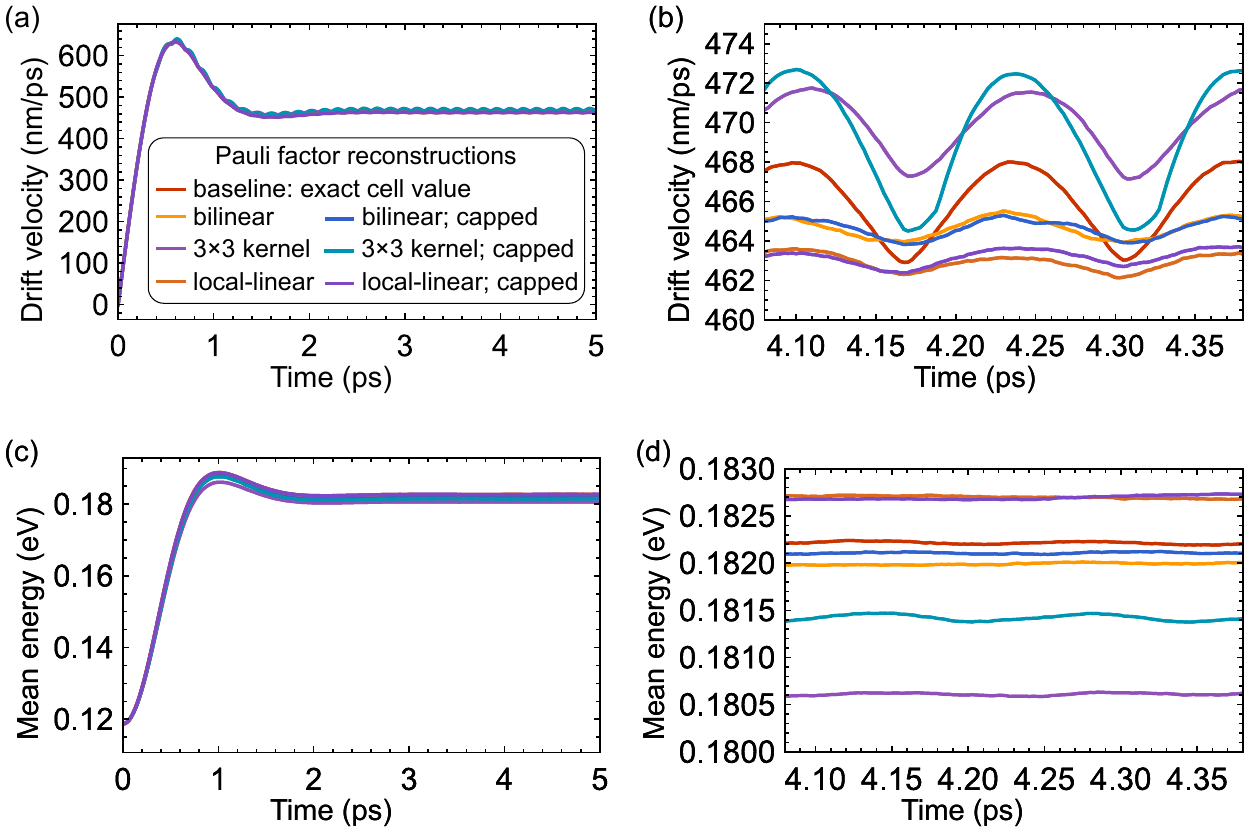}
\caption{Pauli-factor reconstruction tests for the baseline configuration, with $N_p=10^7$ and $N_s=1$. (a) Drift velocity over $0$--$5$ ps. (b) Enlarged steady-state interval. Panels (c) and (d) show the corresponding mean-energy traces. Curve identities are the same in all panels.}
\label{fig:figD2}
\end{figure}

The results presented in Fig.~\ref{fig:figD2} and Table~\ref{tab:tabD2} show that the 
bilinear and local-linear reconstructions substantially reduce the fundamental harmonic. Uncapped reconstructions violate the integer occupancy constraint. The capped variants enforce this constraint but shift the steady-state means, while the capped kernel reconstruction amplifies the harmonic. Taken together, these results support the connection between the oscillation and the cellwise Pauli-factor evaluation but do not establish a bias-free replacement for the exact-cell scheme.

\begin{table}[!th]
\caption{Pauli-factor reconstruction results. The harmonic amplitudes and mean shifts are evaluated over $t\in[3,5]$ ps.}
\label{tab:tabD2}
\centering
\footnotesize
\setlength{\tabcolsep}{2.5pt}
\begin{ruledtabular}
\begin{tabular}{lcccc}
Pauli-factor evaluation & $A/A_0$ &
$\Delta\overline{v}_d$ & $\Delta\overline{\varepsilon}$ & $R_{\max}$ \\
& & (nm/ps) & (meV) & \\
\hline
exact cell value          & 1.00 &  0.00 &  0.000 & 1.000 \\
bilinear                  & 0.27 & -1.39 & -0.208 & 1.099 \\
bilinear; capped          & 0.27 & -1.33 & -0.088 & 1.000 \\
$3\times3$ kernel         & 0.88 & +3.89 & -1.616 & 1.296 \\
$3\times3$ kernel; capped & 1.62 & +3.33 & -0.773 & 1.000 \\
local-linear              & 0.22 & -3.02 & +0.473 & 1.001 \\
local-linear; capped      & 0.22 & -2.94 & +0.488 & 1.000 \\
\end{tabular}
\end{ruledtabular}
\end{table}

\FloatBarrier
\bibliography{refs}

\begin{thebibliography}{35}%
\makeatletter
\providecommand \@ifxundefined [1]{%
 \@ifx{#1\undefined}
}%
\providecommand \@ifnum [1]{%
 \ifnum #1\expandafter \@firstoftwo
 \else \expandafter \@secondoftwo
 \fi
}%
\providecommand \@ifx [1]{%
 \ifx #1\expandafter \@firstoftwo
 \else \expandafter \@secondoftwo
 \fi
}%
\providecommand \natexlab [1]{#1}%
\providecommand \enquote  [1]{``#1''}%
\providecommand \bibnamefont  [1]{#1}%
\providecommand \bibfnamefont [1]{#1}%
\providecommand \citenamefont [1]{#1}%
\providecommand \href@noop [0]{\@secondoftwo}%
\providecommand \href [0]{\begingroup \@sanitize@url \@href}%
\providecommand \@href[1]{\@@startlink{#1}\@@href}%
\providecommand \@@href[1]{\endgroup#1\@@endlink}%
\providecommand \@sanitize@url [0]{\catcode `\\12\catcode `\$12\catcode
  `\&12\catcode `\#12\catcode `\^12\catcode `\_12\catcode `\%12\relax}%
\providecommand \@@startlink[1]{}%
\providecommand \@@endlink[0]{}%
\providecommand \url  [0]{\begingroup\@sanitize@url \@url }%
\providecommand \@url [1]{\endgroup\@href {#1}{\urlprefix }}%
\providecommand \urlprefix  [0]{URL }%
\providecommand \Eprint [0]{\href }%
\providecommand \doibase [0]{https://doi.org/}%
\providecommand \selectlanguage [0]{\@gobble}%
\providecommand \bibinfo  [0]{\@secondoftwo}%
\providecommand \bibfield  [0]{\@secondoftwo}%
\providecommand \translation [1]{[#1]}%
\providecommand \BibitemOpen [0]{}%
\providecommand \bibitemStop [0]{}%
\providecommand \bibitemNoStop [0]{.\EOS\space}%
\providecommand \EOS [0]{\spacefactor3000\relax}%
\providecommand \BibitemShut  [1]{\csname bibitem#1\endcsname}%
\let\auto@bib@innerbib\@empty
\bibitem [{\citenamefont {Metropolis}\ and\ \citenamefont
  {Ulam}(1949)}]{MetropolisUlam1949}%
  \BibitemOpen
  \bibfield  {author} {\bibinfo {author} {\bibfnamefont {N.}~\bibnamefont
  {Metropolis}}\ and\ \bibinfo {author} {\bibfnamefont {S.}~\bibnamefont
  {Ulam}},\ }\bibfield  {title} {\bibinfo {title} {The {M}onte {C}arlo
  method},\ }\href {https://doi.org/10.1080/01621459.1949.10483310} {\bibfield
  {journal} {\bibinfo  {journal} {Journal of the American Statistical
  Association}\ }\textbf {\bibinfo {volume} {44}},\ \bibinfo {pages} {335}
  (\bibinfo {year} {1949})}\BibitemShut {NoStop}%
\bibitem [{\citenamefont {Kurosawa}(1966)}]{kurosawa1966monte}%
  \BibitemOpen
  \bibfield  {author} {\bibinfo {author} {\bibfnamefont {T.}~\bibnamefont
  {Kurosawa}},\ }\bibfield  {title} {\bibinfo {title} {{M}onte {C}arlo
  calculation of hot electron problems},\ }\href@noop {} {\bibfield  {journal}
  {\bibinfo  {journal} {Journal of the Physical Society of Japan}\ }\textbf
  {\bibinfo {volume} {21}},\ \bibinfo {pages} {424} (\bibinfo {year} {1966})},\
  \bibinfo {note} {proceedings of the International Conference on the Physics
  of Semiconductors, Kyoto, 1966}\BibitemShut {NoStop}%
\bibitem [{\citenamefont {Jacoboni}\ and\ \citenamefont
  {Lugli}(1989)}]{JacoboniLugli1989}%
  \BibitemOpen
  \bibfield  {author} {\bibinfo {author} {\bibfnamefont {C.}~\bibnamefont
  {Jacoboni}}\ and\ \bibinfo {author} {\bibfnamefont {P.}~\bibnamefont
  {Lugli}},\ }\href {https://doi.org/10.1007/978-3-7091-6963-6} {\emph
  {\bibinfo {title} {The {M}onte {C}arlo Method for Semiconductor Device
  Simulation}}},\ Computational Microelectronics\ (\bibinfo  {publisher}
  {Springer-Verlag},\ \bibinfo {address} {Wien and New York},\ \bibinfo {year}
  {1989})\BibitemShut {NoStop}%
\bibitem [{\citenamefont {Jacoboni}(2010)}]{Jacoboni2010}%
  \BibitemOpen
  \bibfield  {author} {\bibinfo {author} {\bibfnamefont {C.}~\bibnamefont
  {Jacoboni}},\ }\href {https://doi.org/10.1007/978-3-642-10586-9} {\emph
  {\bibinfo {title} {Theory of Electron Transport in Semiconductors: A Pathway
  from Elementary Physics to Nonequilibrium {G}reen Functions}}},\ \bibinfo
  {series} {Springer Series in Solid-State Sciences}, Vol.\ \bibinfo {volume}
  {165}\ (\bibinfo  {publisher} {Springer},\ \bibinfo {address} {Berlin,
  Heidelberg},\ \bibinfo {year} {2010})\BibitemShut {NoStop}%
\bibitem [{\citenamefont {Novoselov}\ \emph {et~al.}(2004)\citenamefont
  {Novoselov}, \citenamefont {Geim}, \citenamefont {Morozov}, \citenamefont
  {Jiang}, \citenamefont {Zhang}, \citenamefont {Dubonos}, \citenamefont
  {Grigorieva},\ and\ \citenamefont {Firsov}}]{Novoselov2004}%
  \BibitemOpen
  \bibfield  {author} {\bibinfo {author} {\bibfnamefont {K.~S.}\ \bibnamefont
  {Novoselov}}, \bibinfo {author} {\bibfnamefont {A.~K.}\ \bibnamefont {Geim}},
  \bibinfo {author} {\bibfnamefont {S.~V.}\ \bibnamefont {Morozov}}, \bibinfo
  {author} {\bibfnamefont {D.}~\bibnamefont {Jiang}}, \bibinfo {author}
  {\bibfnamefont {Y.}~\bibnamefont {Zhang}}, \bibinfo {author} {\bibfnamefont
  {S.~V.}\ \bibnamefont {Dubonos}}, \bibinfo {author} {\bibfnamefont {I.~V.}\
  \bibnamefont {Grigorieva}},\ and\ \bibinfo {author} {\bibfnamefont {A.~A.}\
  \bibnamefont {Firsov}},\ }\bibfield  {title} {\bibinfo {title} {Electric
  field effect in atomically thin carbon films},\ }\href
  {https://doi.org/10.1126/science.1102896} {\bibfield  {journal} {\bibinfo
  {journal} {Science}\ }\textbf {\bibinfo {volume} {306}},\ \bibinfo {pages}
  {666} (\bibinfo {year} {2004})}\BibitemShut {NoStop}%
\bibitem [{\citenamefont {Castro~Neto}\ \emph {et~al.}(2009)\citenamefont
  {Castro~Neto}, \citenamefont {Guinea}, \citenamefont {Peres}, \citenamefont
  {Novoselov},\ and\ \citenamefont {Geim}}]{CastroNeto2009}%
  \BibitemOpen
  \bibfield  {author} {\bibinfo {author} {\bibfnamefont {A.~H.}\ \bibnamefont
  {Castro~Neto}}, \bibinfo {author} {\bibfnamefont {F.}~\bibnamefont {Guinea}},
  \bibinfo {author} {\bibfnamefont {N.~M.~R.}\ \bibnamefont {Peres}}, \bibinfo
  {author} {\bibfnamefont {K.~S.}\ \bibnamefont {Novoselov}},\ and\ \bibinfo
  {author} {\bibfnamefont {A.~K.}\ \bibnamefont {Geim}},\ }\bibfield  {title}
  {\bibinfo {title} {The electronic properties of graphene},\ }\href
  {https://doi.org/10.1103/RevModPhys.81.109} {\bibfield  {journal} {\bibinfo
  {journal} {Rev. Mod. Phys.}\ }\textbf {\bibinfo {volume} {81}},\ \bibinfo
  {pages} {109} (\bibinfo {year} {2009})}\BibitemShut {NoStop}%
\bibitem [{\citenamefont {Das~Sarma}\ \emph {et~al.}(2011)\citenamefont
  {Das~Sarma}, \citenamefont {Adam}, \citenamefont {Hwang},\ and\ \citenamefont
  {Rossi}}]{DasSarma2011}%
  \BibitemOpen
  \bibfield  {author} {\bibinfo {author} {\bibfnamefont {S.}~\bibnamefont
  {Das~Sarma}}, \bibinfo {author} {\bibfnamefont {S.}~\bibnamefont {Adam}},
  \bibinfo {author} {\bibfnamefont {E.~H.}\ \bibnamefont {Hwang}},\ and\
  \bibinfo {author} {\bibfnamefont {E.}~\bibnamefont {Rossi}},\ }\bibfield
  {title} {\bibinfo {title} {Electronic transport in two-dimensional
  graphene},\ }\href {https://doi.org/10.1103/RevModPhys.83.407} {\bibfield
  {journal} {\bibinfo  {journal} {Rev. Mod. Phys.}\ }\textbf {\bibinfo {volume}
  {83}},\ \bibinfo {pages} {407} (\bibinfo {year} {2011})}\BibitemShut
  {NoStop}%
\bibitem [{\citenamefont {Coco}\ \emph {et~al.}(2017)\citenamefont {Coco},
  \citenamefont {Majorana},\ and\ \citenamefont {Romano}}]{Coco2017}%
  \BibitemOpen
  \bibfield  {author} {\bibinfo {author} {\bibfnamefont {M.}~\bibnamefont
  {Coco}}, \bibinfo {author} {\bibfnamefont {A.}~\bibnamefont {Majorana}},\
  and\ \bibinfo {author} {\bibfnamefont {V.}~\bibnamefont {Romano}},\
  }\bibfield  {title} {\bibinfo {title} {Cross validation of discontinuous
  {G}alerkin method and {M}onte {C}arlo simulations of charge transport in
  graphene on substrate},\ }\href {https://doi.org/10.1007/s11587-016-0298-4}
  {\bibfield  {journal} {\bibinfo  {journal} {Ricerche di Matematica}\ }\textbf
  {\bibinfo {volume} {66}},\ \bibinfo {pages} {201} (\bibinfo {year}
  {2017})}\BibitemShut {NoStop}%
\bibitem [{\citenamefont {Majorana}\ \emph {et~al.}(2019)\citenamefont
  {Majorana}, \citenamefont {Nastasi},\ and\ \citenamefont
  {Romano}}]{MajoranaNastasiRomano2019}%
  \BibitemOpen
  \bibfield  {author} {\bibinfo {author} {\bibfnamefont {A.}~\bibnamefont
  {Majorana}}, \bibinfo {author} {\bibfnamefont {G.}~\bibnamefont {Nastasi}},\
  and\ \bibinfo {author} {\bibfnamefont {V.}~\bibnamefont {Romano}},\
  }\bibfield  {title} {\bibinfo {title} {Simulation of bipolar charge transport
  in graphene by using a discontinuous {G}alerkin method},\ }\href
  {https://doi.org/10.4208/cicp.OA-2018-0052} {\bibfield  {journal} {\bibinfo
  {journal} {Communications in Computational Physics}\ }\textbf {\bibinfo
  {volume} {26}},\ \bibinfo {pages} {114} (\bibinfo {year} {2019})}\BibitemShut
  {NoStop}%
\bibitem [{\citenamefont {Li}\ \emph {et~al.}(2010)\citenamefont {Li},
  \citenamefont {Barry}, \citenamefont {Zavada}, \citenamefont
  {Buongiorno~Nardelli},\ and\ \citenamefont {Kim}}]{Li2010}%
  \BibitemOpen
  \bibfield  {author} {\bibinfo {author} {\bibfnamefont {X.}~\bibnamefont
  {Li}}, \bibinfo {author} {\bibfnamefont {E.~A.}\ \bibnamefont {Barry}},
  \bibinfo {author} {\bibfnamefont {J.~M.}\ \bibnamefont {Zavada}}, \bibinfo
  {author} {\bibfnamefont {M.}~\bibnamefont {Buongiorno~Nardelli}},\ and\
  \bibinfo {author} {\bibfnamefont {K.~W.}\ \bibnamefont {Kim}},\ }\bibfield
  {title} {\bibinfo {title} {Influence of electron-electron scattering on
  transport characteristics in monolayer graphene},\ }\href
  {https://doi.org/10.1063/1.3483612} {\bibfield  {journal} {\bibinfo
  {journal} {Applied Physics Letters}\ }\textbf {\bibinfo {volume} {97}},\
  \bibinfo {pages} {082101} (\bibinfo {year} {2010})}\BibitemShut {NoStop}%
\bibitem [{\citenamefont {Fang}\ \emph {et~al.}(2011)\citenamefont {Fang},
  \citenamefont {Konar}, \citenamefont {Xing},\ and\ \citenamefont
  {Jena}}]{Fang2011}%
  \BibitemOpen
  \bibfield  {author} {\bibinfo {author} {\bibfnamefont {T.}~\bibnamefont
  {Fang}}, \bibinfo {author} {\bibfnamefont {A.}~\bibnamefont {Konar}},
  \bibinfo {author} {\bibfnamefont {H.}~\bibnamefont {Xing}},\ and\ \bibinfo
  {author} {\bibfnamefont {D.}~\bibnamefont {Jena}},\ }\bibfield  {title}
  {\bibinfo {title} {High-field transport in two-dimensional graphene},\ }\href
  {https://doi.org/10.1103/PhysRevB.84.125450} {\bibfield  {journal} {\bibinfo
  {journal} {Phys. Rev. B}\ }\textbf {\bibinfo {volume} {84}},\ \bibinfo
  {pages} {125450} (\bibinfo {year} {2011})}\BibitemShut {NoStop}%
\bibitem [{\citenamefont {Romano}\ \emph {et~al.}(2015)\citenamefont {Romano},
  \citenamefont {Majorana},\ and\ \citenamefont {Coco}}]{Romano2015}%
  \BibitemOpen
  \bibfield  {author} {\bibinfo {author} {\bibfnamefont {V.}~\bibnamefont
  {Romano}}, \bibinfo {author} {\bibfnamefont {A.}~\bibnamefont {Majorana}},\
  and\ \bibinfo {author} {\bibfnamefont {M.}~\bibnamefont {Coco}},\ }\bibfield
  {title} {\bibinfo {title} {{DSMC} method consistent with the {P}auli
  exclusion principle and comparison with deterministic solutions for charge
  transport in graphene},\ }\href {https://doi.org/10.1016/j.jcp.2015.08.047}
  {\bibfield  {journal} {\bibinfo  {journal} {Journal of Computational
  Physics}\ }\textbf {\bibinfo {volume} {302}},\ \bibinfo {pages} {267}
  (\bibinfo {year} {2015})}\BibitemShut {NoStop}%
\bibitem [{\citenamefont {Nastasi}\ and\ \citenamefont
  {Romano}(2023)}]{Nastasi2022}%
  \BibitemOpen
  \bibfield  {author} {\bibinfo {author} {\bibfnamefont {G.}~\bibnamefont
  {Nastasi}}\ and\ \bibinfo {author} {\bibfnamefont {V.}~\bibnamefont
  {Romano}},\ }\bibfield  {title} {\bibinfo {title} {Mathematical aspects and
  simulation of electron--electron scattering in graphene},\ }\href
  {https://doi.org/10.1007/s00033-022-01912-8} {\bibfield  {journal} {\bibinfo
  {journal} {Zeitschrift f{\"u}r angewandte Mathematik und Physik}\ }\textbf
  {\bibinfo {volume} {74}},\ \bibinfo {pages} {28} (\bibinfo {year}
  {2023})}\BibitemShut {NoStop}%
\bibitem [{\citenamefont {Lugli}\ and\ \citenamefont
  {Ferry}(1985)}]{LugliFerry1985}%
  \BibitemOpen
  \bibfield  {author} {\bibinfo {author} {\bibfnamefont {P.}~\bibnamefont
  {Lugli}}\ and\ \bibinfo {author} {\bibfnamefont {D.~K.}\ \bibnamefont
  {Ferry}},\ }\bibfield  {title} {\bibinfo {title} {Degeneracy in the ensemble
  {M}onte {C}arlo method for high-field transport in semiconductors},\ }\href
  {https://doi.org/10.1109/T-ED.1985.22291} {\bibfield  {journal} {\bibinfo
  {journal} {IEEE Transactions on Electron Devices}\ }\textbf {\bibinfo
  {volume} {32}},\ \bibinfo {pages} {2431} (\bibinfo {year}
  {1985})}\BibitemShut {NoStop}%
\bibitem [{\citenamefont {Tadyszak}\ \emph {et~al.}(1996)\citenamefont
  {Tadyszak}, \citenamefont {Danneville}, \citenamefont {Cappy}, \citenamefont
  {Reggiani}, \citenamefont {Varani},\ and\ \citenamefont
  {Rota}}]{Tadyszak1996}%
  \BibitemOpen
  \bibfield  {author} {\bibinfo {author} {\bibfnamefont {P.}~\bibnamefont
  {Tadyszak}}, \bibinfo {author} {\bibfnamefont {F.}~\bibnamefont
  {Danneville}}, \bibinfo {author} {\bibfnamefont {A.}~\bibnamefont {Cappy}},
  \bibinfo {author} {\bibfnamefont {L.}~\bibnamefont {Reggiani}}, \bibinfo
  {author} {\bibfnamefont {L.}~\bibnamefont {Varani}},\ and\ \bibinfo {author}
  {\bibfnamefont {L.}~\bibnamefont {Rota}},\ }\bibfield  {title} {\bibinfo
  {title} {{M}onte {C}arlo calculations of hot-carrier noise under degenerate
  conditions},\ }\href {https://doi.org/10.1063/1.117611} {\bibfield  {journal}
  {\bibinfo  {journal} {Applied Physics Letters}\ }\textbf {\bibinfo {volume}
  {69}},\ \bibinfo {pages} {1450} (\bibinfo {year} {1996})}\BibitemShut
  {NoStop}%
\bibitem [{\citenamefont {Borowik}\ and\ \citenamefont
  {Thobel}(1998)}]{BorowikThobel1998}%
  \BibitemOpen
  \bibfield  {author} {\bibinfo {author} {\bibfnamefont {P.}~\bibnamefont
  {Borowik}}\ and\ \bibinfo {author} {\bibfnamefont {J.-L.}\ \bibnamefont
  {Thobel}},\ }\bibfield  {title} {\bibinfo {title} {Improved {M}onte {C}arlo
  method for the study of electron transport in degenerate semiconductors},\
  }\href {https://doi.org/10.1063/1.368547} {\bibfield  {journal} {\bibinfo
  {journal} {Journal of Applied Physics}\ }\textbf {\bibinfo {volume} {84}},\
  \bibinfo {pages} {3706} (\bibinfo {year} {1998})}\BibitemShut {NoStop}%
\bibitem [{\citenamefont {Borowik}\ and\ \citenamefont
  {Adamowicz}(2005)}]{BorowikAdamowicz2005}%
  \BibitemOpen
  \bibfield  {author} {\bibinfo {author} {\bibfnamefont {P.}~\bibnamefont
  {Borowik}}\ and\ \bibinfo {author} {\bibfnamefont {L.}~\bibnamefont
  {Adamowicz}},\ }\bibfield  {title} {\bibinfo {title} {Improved algorithm for
  {M}onte {C}arlo studies of electron transport in degenerate semiconductors},\
  }\href {https://doi.org/10.1016/j.physb.2005.05.021} {\bibfield  {journal}
  {\bibinfo  {journal} {Physica B: Condensed Matter}\ }\textbf {\bibinfo
  {volume} {365}},\ \bibinfo {pages} {235} (\bibinfo {year}
  {2005})}\BibitemShut {NoStop}%
\bibitem [{\citenamefont {Coco}\ \emph {et~al.}(2021)\citenamefont {Coco},
  \citenamefont {Bordone}, \citenamefont {Demeio},\ and\ \citenamefont
  {Romano}}]{Coco2021}%
  \BibitemOpen
  \bibfield  {author} {\bibinfo {author} {\bibfnamefont {M.}~\bibnamefont
  {Coco}}, \bibinfo {author} {\bibfnamefont {P.}~\bibnamefont {Bordone}},
  \bibinfo {author} {\bibfnamefont {L.}~\bibnamefont {Demeio}},\ and\ \bibinfo
  {author} {\bibfnamefont {V.}~\bibnamefont {Romano}},\ }\bibfield  {title}
  {\bibinfo {title} {Pauli principle and the {M}onte {C}arlo method for charge
  transport in graphene},\ }\href {https://doi.org/10.1103/PhysRevB.104.205410}
  {\bibfield  {journal} {\bibinfo  {journal} {Phys. Rev. B}\ }\textbf {\bibinfo
  {volume} {104}},\ \bibinfo {pages} {205410} (\bibinfo {year}
  {2021})}\BibitemShut {NoStop}%
\bibitem [{\citenamefont {Kotov}\ \emph {et~al.}(2012)\citenamefont {Kotov},
  \citenamefont {Uchoa}, \citenamefont {Pereira}, \citenamefont {Guinea},\ and\
  \citenamefont {Castro~Neto}}]{Kotov2012}%
  \BibitemOpen
  \bibfield  {author} {\bibinfo {author} {\bibfnamefont {V.~N.}\ \bibnamefont
  {Kotov}}, \bibinfo {author} {\bibfnamefont {B.}~\bibnamefont {Uchoa}},
  \bibinfo {author} {\bibfnamefont {V.~M.}\ \bibnamefont {Pereira}}, \bibinfo
  {author} {\bibfnamefont {F.}~\bibnamefont {Guinea}},\ and\ \bibinfo {author}
  {\bibfnamefont {A.~H.}\ \bibnamefont {Castro~Neto}},\ }\bibfield  {title}
  {\bibinfo {title} {Electron-electron interactions in graphene: Current status
  and perspectives},\ }\href {https://doi.org/10.1103/RevModPhys.84.1067}
  {\bibfield  {journal} {\bibinfo  {journal} {Rev. Mod. Phys.}\ }\textbf
  {\bibinfo {volume} {84}},\ \bibinfo {pages} {1067} (\bibinfo {year}
  {2012})}\BibitemShut {NoStop}%
\bibitem [{\citenamefont {Dawlaty}\ \emph {et~al.}(2008)\citenamefont
  {Dawlaty}, \citenamefont {Shivaraman}, \citenamefont {Chandrashekhar},
  \citenamefont {Rana},\ and\ \citenamefont {Spencer}}]{Dawlaty2008}%
  \BibitemOpen
  \bibfield  {author} {\bibinfo {author} {\bibfnamefont {J.~M.}\ \bibnamefont
  {Dawlaty}}, \bibinfo {author} {\bibfnamefont {S.}~\bibnamefont {Shivaraman}},
  \bibinfo {author} {\bibfnamefont {M.}~\bibnamefont {Chandrashekhar}},
  \bibinfo {author} {\bibfnamefont {F.}~\bibnamefont {Rana}},\ and\ \bibinfo
  {author} {\bibfnamefont {M.~G.}\ \bibnamefont {Spencer}},\ }\bibfield
  {title} {\bibinfo {title} {Measurement of ultrafast carrier dynamics in
  epitaxial graphene},\ }\href {https://doi.org/10.1063/1.2837539} {\bibfield
  {journal} {\bibinfo  {journal} {Applied Physics Letters}\ }\textbf {\bibinfo
  {volume} {92}},\ \bibinfo {pages} {042116} (\bibinfo {year}
  {2008})}\BibitemShut {NoStop}%
\bibitem [{\citenamefont {Kim}\ \emph {et~al.}(2020)\citenamefont {Kim},
  \citenamefont {Xu}, \citenamefont {Berdyugin}, \citenamefont {Principi},
  \citenamefont {Slizovskiy}, \citenamefont {Xin}, \citenamefont
  {Kumaravadivel}, \citenamefont {Kuang}, \citenamefont {Hamer}, \citenamefont
  {Krishna~Kumar}, \citenamefont {Gorbachev}, \citenamefont {Watanabe},
  \citenamefont {Taniguchi}, \citenamefont {Grigorieva}, \citenamefont
  {Fal'ko}, \citenamefont {Polini},\ and\ \citenamefont {Geim}}]{Kim2020}%
  \BibitemOpen
  \bibfield  {author} {\bibinfo {author} {\bibfnamefont {M.}~\bibnamefont
  {Kim}}, \bibinfo {author} {\bibfnamefont {S.~G.}\ \bibnamefont {Xu}},
  \bibinfo {author} {\bibfnamefont {A.~I.}\ \bibnamefont {Berdyugin}}, \bibinfo
  {author} {\bibfnamefont {A.}~\bibnamefont {Principi}}, \bibinfo {author}
  {\bibfnamefont {S.}~\bibnamefont {Slizovskiy}}, \bibinfo {author}
  {\bibfnamefont {N.}~\bibnamefont {Xin}}, \bibinfo {author} {\bibfnamefont
  {P.}~\bibnamefont {Kumaravadivel}}, \bibinfo {author} {\bibfnamefont
  {W.}~\bibnamefont {Kuang}}, \bibinfo {author} {\bibfnamefont
  {M.}~\bibnamefont {Hamer}}, \bibinfo {author} {\bibfnamefont
  {R.}~\bibnamefont {Krishna~Kumar}}, \bibinfo {author} {\bibfnamefont {R.~V.}\
  \bibnamefont {Gorbachev}}, \bibinfo {author} {\bibfnamefont {K.}~\bibnamefont
  {Watanabe}}, \bibinfo {author} {\bibfnamefont {T.}~\bibnamefont {Taniguchi}},
  \bibinfo {author} {\bibfnamefont {I.~V.}\ \bibnamefont {Grigorieva}},
  \bibinfo {author} {\bibfnamefont {V.~I.}\ \bibnamefont {Fal'ko}}, \bibinfo
  {author} {\bibfnamefont {M.}~\bibnamefont {Polini}},\ and\ \bibinfo {author}
  {\bibfnamefont {A.~K.}\ \bibnamefont {Geim}},\ }\bibfield  {title} {\bibinfo
  {title} {Control of electron-electron interaction in graphene by proximity
  screening},\ }\href {https://doi.org/10.1038/s41467-020-15829-1} {\bibfield
  {journal} {\bibinfo  {journal} {Nature Communications}\ }\textbf {\bibinfo
  {volume} {11}},\ \bibinfo {pages} {2339} (\bibinfo {year}
  {2020})}\BibitemShut {NoStop}%
\bibitem [{\citenamefont {Hwang}\ and\ \citenamefont
  {Das~Sarma}(2007)}]{Hwang2007}%
  \BibitemOpen
  \bibfield  {author} {\bibinfo {author} {\bibfnamefont {E.~H.}\ \bibnamefont
  {Hwang}}\ and\ \bibinfo {author} {\bibfnamefont {S.}~\bibnamefont
  {Das~Sarma}},\ }\bibfield  {title} {\bibinfo {title} {Dielectric function,
  screening, and plasmons in two-dimensional graphene},\ }\href
  {https://doi.org/10.1103/PhysRevB.75.205418} {\bibfield  {journal} {\bibinfo
  {journal} {Phys. Rev. B}\ }\textbf {\bibinfo {volume} {75}},\ \bibinfo
  {pages} {205418} (\bibinfo {year} {2007})}\BibitemShut {NoStop}%
\bibitem [{\citenamefont {Brunetti}\ \emph {et~al.}(1985)\citenamefont
  {Brunetti}, \citenamefont {Jacoboni}, \citenamefont {Matulionis},\ and\
  \citenamefont {Dienys}}]{Brunetti1985}%
  \BibitemOpen
  \bibfield  {author} {\bibinfo {author} {\bibfnamefont {R.}~\bibnamefont
  {Brunetti}}, \bibinfo {author} {\bibfnamefont {C.}~\bibnamefont {Jacoboni}},
  \bibinfo {author} {\bibfnamefont {A.}~\bibnamefont {Matulionis}},\ and\
  \bibinfo {author} {\bibfnamefont {V.}~\bibnamefont {Dienys}},\ }\bibfield
  {title} {\bibinfo {title} {Effect of interparticle collisions on energy
  relaxation of carriers in semiconductors},\ }\href
  {https://doi.org/10.1016/0378-4363(85)90373-0} {\bibfield  {journal}
  {\bibinfo  {journal} {Physica B+C}\ }\textbf {\bibinfo {volume} {134}},\
  \bibinfo {pages} {369} (\bibinfo {year} {1985})}\BibitemShut {NoStop}%
\bibitem [{\citenamefont {Goodnick}\ and\ \citenamefont
  {Lugli}(1988)}]{GoodnickLugli1988}%
  \BibitemOpen
  \bibfield  {author} {\bibinfo {author} {\bibfnamefont {S.~M.}\ \bibnamefont
  {Goodnick}}\ and\ \bibinfo {author} {\bibfnamefont {P.}~\bibnamefont
  {Lugli}},\ }\bibfield  {title} {\bibinfo {title} {Effect of electron-electron
  scattering on nonequilibrium transport in quantum-well systems},\ }\href
  {https://doi.org/10.1103/PhysRevB.37.2578} {\bibfield  {journal} {\bibinfo
  {journal} {Phys. Rev. B}\ }\textbf {\bibinfo {volume} {37}},\ \bibinfo
  {pages} {2578} (\bibinfo {year} {1988})}\BibitemShut {NoStop}%
\bibitem [{\citenamefont {Mo{\v{s}}ko}\ and\ \citenamefont
  {Mo{\v{s}}kov{\'a}}(1991)}]{MoskoMoskova1991}%
  \BibitemOpen
  \bibfield  {author} {\bibinfo {author} {\bibfnamefont {M.}~\bibnamefont
  {Mo{\v{s}}ko}}\ and\ \bibinfo {author} {\bibfnamefont {A.}~\bibnamefont
  {Mo{\v{s}}kov{\'a}}},\ }\bibfield  {title} {\bibinfo {title} {Ensemble
  {M}onte {C}arlo simulation of electron-electron scattering: Improvements of
  conventional methods},\ }\href {https://doi.org/10.1103/PhysRevB.44.10794}
  {\bibfield  {journal} {\bibinfo  {journal} {Phys. Rev. B}\ }\textbf {\bibinfo
  {volume} {44}},\ \bibinfo {pages} {10794} (\bibinfo {year}
  {1991})}\BibitemShut {NoStop}%
\bibitem [{\citenamefont {Coco}\ and\ \citenamefont
  {Romano}(2018)}]{CocoRomano2018}%
  \BibitemOpen
  \bibfield  {author} {\bibinfo {author} {\bibfnamefont {M.}~\bibnamefont
  {Coco}}\ and\ \bibinfo {author} {\bibfnamefont {V.}~\bibnamefont {Romano}},\
  }\bibfield  {title} {\bibinfo {title} {Assessment of the constant phonon
  relaxation time approximation in electron--phonon coupling in graphene},\
  }\href {https://doi.org/10.1080/23324309.2018.1558253} {\bibfield  {journal}
  {\bibinfo  {journal} {Journal of Computational and Theoretical Transport}\
  }\textbf {\bibinfo {volume} {47}},\ \bibinfo {pages} {246} (\bibinfo {year}
  {2018})}\BibitemShut {NoStop}%
\bibitem [{\citenamefont {Kim}\ \emph {et~al.}(2015)\citenamefont {Kim},
  \citenamefont {Kim}, \citenamefont {Cho}, \citenamefont {Ryoo}, \citenamefont
  {Park}, \citenamefont {Kim}, \citenamefont {Kim}, \citenamefont {Lee},
  \citenamefont {Li}, \citenamefont {Park}, \citenamefont {Yoo}, \citenamefont
  {Yoon}, \citenamefont {Dorgan}, \citenamefont {Pop}, \citenamefont {Heinz},
  \citenamefont {Hone}, \citenamefont {Chun}, \citenamefont {Cheong},
  \citenamefont {Lee}, \citenamefont {Bae},\ and\ \citenamefont
  {Park}}]{Kim2015}%
  \BibitemOpen
  \bibfield  {author} {\bibinfo {author} {\bibfnamefont {Y.~D.}\ \bibnamefont
  {Kim}}, \bibinfo {author} {\bibfnamefont {H.}~\bibnamefont {Kim}}, \bibinfo
  {author} {\bibfnamefont {Y.}~\bibnamefont {Cho}}, \bibinfo {author}
  {\bibfnamefont {J.~H.}\ \bibnamefont {Ryoo}}, \bibinfo {author}
  {\bibfnamefont {C.-H.}\ \bibnamefont {Park}}, \bibinfo {author}
  {\bibfnamefont {P.}~\bibnamefont {Kim}}, \bibinfo {author} {\bibfnamefont
  {Y.~S.}\ \bibnamefont {Kim}}, \bibinfo {author} {\bibfnamefont
  {S.}~\bibnamefont {Lee}}, \bibinfo {author} {\bibfnamefont {Y.}~\bibnamefont
  {Li}}, \bibinfo {author} {\bibfnamefont {S.-N.}\ \bibnamefont {Park}},
  \bibinfo {author} {\bibfnamefont {Y.~S.}\ \bibnamefont {Yoo}}, \bibinfo
  {author} {\bibfnamefont {D.}~\bibnamefont {Yoon}}, \bibinfo {author}
  {\bibfnamefont {V.~E.}\ \bibnamefont {Dorgan}}, \bibinfo {author}
  {\bibfnamefont {E.}~\bibnamefont {Pop}}, \bibinfo {author} {\bibfnamefont
  {T.~F.}\ \bibnamefont {Heinz}}, \bibinfo {author} {\bibfnamefont
  {J.}~\bibnamefont {Hone}}, \bibinfo {author} {\bibfnamefont {S.-H.}\
  \bibnamefont {Chun}}, \bibinfo {author} {\bibfnamefont {H.}~\bibnamefont
  {Cheong}}, \bibinfo {author} {\bibfnamefont {S.~W.}\ \bibnamefont {Lee}},
  \bibinfo {author} {\bibfnamefont {M.-H.}\ \bibnamefont {Bae}},\ and\ \bibinfo
  {author} {\bibfnamefont {Y.~D.}\ \bibnamefont {Park}},\ }\bibfield  {title}
  {\bibinfo {title} {Bright visible light emission from graphene},\ }\href
  {https://doi.org/10.1038/nnano.2015.118} {\bibfield  {journal} {\bibinfo
  {journal} {Nature Nanotechnology}\ }\textbf {\bibinfo {volume} {10}},\
  \bibinfo {pages} {676} (\bibinfo {year} {2015})}\BibitemShut {NoStop}%
\bibitem [{\citenamefont {Reed}\ \emph {et~al.}(2010)\citenamefont {Reed},
  \citenamefont {Uchoa}, \citenamefont {Joe}, \citenamefont {Gan},
  \citenamefont {Casa}, \citenamefont {Fradkin},\ and\ \citenamefont
  {Abbamonte}}]{Reed2010}%
  \BibitemOpen
  \bibfield  {author} {\bibinfo {author} {\bibfnamefont {J.~P.}\ \bibnamefont
  {Reed}}, \bibinfo {author} {\bibfnamefont {B.}~\bibnamefont {Uchoa}},
  \bibinfo {author} {\bibfnamefont {Y.~I.}\ \bibnamefont {Joe}}, \bibinfo
  {author} {\bibfnamefont {Y.}~\bibnamefont {Gan}}, \bibinfo {author}
  {\bibfnamefont {D.}~\bibnamefont {Casa}}, \bibinfo {author} {\bibfnamefont
  {E.}~\bibnamefont {Fradkin}},\ and\ \bibinfo {author} {\bibfnamefont
  {P.}~\bibnamefont {Abbamonte}},\ }\bibfield  {title} {\bibinfo {title} {The
  effective fine-structure constant of freestanding graphene measured in
  graphite},\ }\href {https://doi.org/10.1126/science.1190920} {\bibfield
  {journal} {\bibinfo  {journal} {Science}\ }\textbf {\bibinfo {volume}
  {330}},\ \bibinfo {pages} {805} (\bibinfo {year} {2010})}\BibitemShut
  {NoStop}%
\bibitem [{\citenamefont {Sabbaghi}\ \emph {et~al.}(2015)\citenamefont
  {Sabbaghi}, \citenamefont {Lee}, \citenamefont {Stauber},\ and\ \citenamefont
  {Kim}}]{Sabbaghi2015}%
  \BibitemOpen
  \bibfield  {author} {\bibinfo {author} {\bibfnamefont {M.}~\bibnamefont
  {Sabbaghi}}, \bibinfo {author} {\bibfnamefont {H.-W.}\ \bibnamefont {Lee}},
  \bibinfo {author} {\bibfnamefont {T.}~\bibnamefont {Stauber}},\ and\ \bibinfo
  {author} {\bibfnamefont {K.~S.}\ \bibnamefont {Kim}},\ }\bibfield  {title}
  {\bibinfo {title} {Drift-induced modifications to the dynamical polarization
  of graphene},\ }\href {https://doi.org/10.1103/PhysRevB.92.195429} {\bibfield
   {journal} {\bibinfo  {journal} {Phys. Rev. B}\ }\textbf {\bibinfo {volume}
  {92}},\ \bibinfo {pages} {195429} (\bibinfo {year} {2015})}\BibitemShut
  {NoStop}%
\bibitem [{\citenamefont {Nastasi}\ and\ \citenamefont
  {Romano}(2020)}]{NastasiRomano2020GFET}%
  \BibitemOpen
  \bibfield  {author} {\bibinfo {author} {\bibfnamefont {G.}~\bibnamefont
  {Nastasi}}\ and\ \bibinfo {author} {\bibfnamefont {V.}~\bibnamefont
  {Romano}},\ }\bibfield  {title} {\bibinfo {title} {A full coupled
  drift-diffusion-{P}oisson simulation of a {GFET}},\ }\href
  {https://doi.org/10.1016/j.cnsns.2020.105300} {\bibfield  {journal} {\bibinfo
   {journal} {Communications in Nonlinear Science and Numerical Simulation}\
  }\textbf {\bibinfo {volume} {87}},\ \bibinfo {pages} {105300} (\bibinfo
  {year} {2020})}\BibitemShut {NoStop}%
\bibitem [{\citenamefont {Nastasi}\ and\ \citenamefont
  {Romano}(2021)}]{NastasiRomano2021EfficientGFET}%
  \BibitemOpen
  \bibfield  {author} {\bibinfo {author} {\bibfnamefont {G.}~\bibnamefont
  {Nastasi}}\ and\ \bibinfo {author} {\bibfnamefont {V.}~\bibnamefont
  {Romano}},\ }\bibfield  {title} {\bibinfo {title} {An efficient {GFET}
  structure},\ }\href {https://doi.org/10.1109/TED.2021.3096492} {\bibfield
  {journal} {\bibinfo  {journal} {IEEE Transactions on Electron Devices}\
  }\textbf {\bibinfo {volume} {68}},\ \bibinfo {pages} {4729} (\bibinfo {year}
  {2021})}\BibitemShut {NoStop}%
\bibitem [{\citenamefont {Singh}\ and\ \citenamefont
  {Battiato}(2020)}]{SinghBattiato2020}%
  \BibitemOpen
  \bibfield  {author} {\bibinfo {author} {\bibfnamefont {S.}~\bibnamefont
  {Singh}}\ and\ \bibinfo {author} {\bibfnamefont {M.}~\bibnamefont
  {Battiato}},\ }\bibfield  {title} {\bibinfo {title} {Effect of strong
  electric fields on material responses: The {B}loch oscillation resonance in
  high field conductivities},\ }\href {https://doi.org/10.3390/ma13051070}
  {\bibfield  {journal} {\bibinfo  {journal} {Materials}\ }\textbf {\bibinfo
  {volume} {13}},\ \bibinfo {pages} {1070} (\bibinfo {year}
  {2020})}\BibitemShut {NoStop}%
\bibitem [{\citenamefont {Cheng}\ \emph {et~al.}(2013)\citenamefont {Cheng},
  \citenamefont {Gamba},\ and\ \citenamefont
  {Morrison}}]{ChengGambaMorrison2013}%
  \BibitemOpen
  \bibfield  {author} {\bibinfo {author} {\bibfnamefont {Y.}~\bibnamefont
  {Cheng}}, \bibinfo {author} {\bibfnamefont {I.~M.}\ \bibnamefont {Gamba}},\
  and\ \bibinfo {author} {\bibfnamefont {P.~J.}\ \bibnamefont {Morrison}},\
  }\bibfield  {title} {\bibinfo {title} {Study of conservation and recurrence
  of {R}unge--{K}utta discontinuous {G}alerkin schemes for {V}lasov--{P}oisson
  systems},\ }\href {https://doi.org/10.1007/s10915-012-9680-x} {\bibfield
  {journal} {\bibinfo  {journal} {Journal of Scientific Computing}\ }\textbf
  {\bibinfo {volume} {56}},\ \bibinfo {pages} {319} (\bibinfo {year}
  {2013})}\BibitemShut {NoStop}%
\bibitem [{\citenamefont {Borysenko}\ \emph {et~al.}(2010)\citenamefont
  {Borysenko}, \citenamefont {Mullen}, \citenamefont {Barry}, \citenamefont
  {Paul}, \citenamefont {Semenov}, \citenamefont {Zavada}, \citenamefont
  {Buongiorno~Nardelli},\ and\ \citenamefont {Kim}}]{Borysenko2010}%
  \BibitemOpen
  \bibfield  {author} {\bibinfo {author} {\bibfnamefont {K.~M.}\ \bibnamefont
  {Borysenko}}, \bibinfo {author} {\bibfnamefont {J.~T.}\ \bibnamefont
  {Mullen}}, \bibinfo {author} {\bibfnamefont {E.~A.}\ \bibnamefont {Barry}},
  \bibinfo {author} {\bibfnamefont {S.}~\bibnamefont {Paul}}, \bibinfo {author}
  {\bibfnamefont {Y.~G.}\ \bibnamefont {Semenov}}, \bibinfo {author}
  {\bibfnamefont {J.~M.}\ \bibnamefont {Zavada}}, \bibinfo {author}
  {\bibfnamefont {M.}~\bibnamefont {Buongiorno~Nardelli}},\ and\ \bibinfo
  {author} {\bibfnamefont {K.~W.}\ \bibnamefont {Kim}},\ }\bibfield  {title}
  {\bibinfo {title} {First-principles analysis of electron-phonon interactions
  in graphene},\ }\href {https://doi.org/10.1103/PhysRevB.81.121412} {\bibfield
   {journal} {\bibinfo  {journal} {Phys. Rev. B}\ }\textbf {\bibinfo {volume}
  {81}},\ \bibinfo {pages} {121412} (\bibinfo {year} {2010})}\BibitemShut
  {NoStop}%
\bibitem [{\citenamefont {Kaasbjerg}\ \emph {et~al.}(2012)\citenamefont
  {Kaasbjerg}, \citenamefont {Thygesen},\ and\ \citenamefont
  {Jacobsen}}]{Kaasbjerg2012}%
  \BibitemOpen
  \bibfield  {author} {\bibinfo {author} {\bibfnamefont {K.}~\bibnamefont
  {Kaasbjerg}}, \bibinfo {author} {\bibfnamefont {K.~S.}\ \bibnamefont
  {Thygesen}},\ and\ \bibinfo {author} {\bibfnamefont {K.~W.}\ \bibnamefont
  {Jacobsen}},\ }\bibfield  {title} {\bibinfo {title} {Unraveling the acoustic
  electron-phonon interaction in graphene},\ }\href
  {https://doi.org/10.1103/PhysRevB.85.165440} {\bibfield  {journal} {\bibinfo
  {journal} {Phys. Rev. B}\ }\textbf {\bibinfo {volume} {85}},\ \bibinfo
  {pages} {165440} (\bibinfo {year} {2012})}\BibitemShut {NoStop}%
\end{thebibliography}%
\end{document}